\documentclass[aps, twocolumn]{revtex4-2}

\usepackage{epsfig}
\usepackage{graphicx}
\usepackage{amsmath,amssymb,amsfonts}
\usepackage{array}
\usepackage{url}
\usepackage[usenames,dvipsnames]{color}
\usepackage[colorlinks=true,linkcolor=magenta, citecolor = blue]{hyperref}
\usepackage{multirow}
\usepackage{float}
\usepackage{lineno}
\usepackage{xspace}
 \usepackage{subcaption}
\usepackage{ulem}
\usepackage{lipsum} 
\usepackage{bigints}

\begin{document}


\title{Thermal Evolution of Magnetars under $f(R, T)$ Gravity}

\author{Charul Rathod}
\email{charulrathod1813@gmail.com}
\author{M. Mishra}
\affiliation{Department of Physics, Birla Institute of Technology and Science, Pilani Campus, Rajasthan, India}
\author{Prasanta Kumar Das}
\affiliation{Department of Physics, Birla Institute of Technology and Science, K. K. Birla Goa Campus, NH-17B, Zuarinagar, Sancoale, Goa 403726, India}
\author{Captain R. Singh}
\affiliation{Department of Physics, Indian Institute of Technology Indore, Simrol, Indore-453552, India}

\begin{abstract}
The present study explores the thermal evolution and emission properties of neutron stars within the framework of modified $f(R, T)$ gravity by solving the coupled energy-balance and heat-transport equations. We compute stellar mass and pressure profiles by solving the Tolman-Oppenheimer-Volkoff equations in both Einstein gravity and modified gravity, employing the APR, FPS, and SLy equations of state, with and without the strong magnetic field. Using these profiles, we assess the red-shifted surface temperature, $T_s^{\infty}$,  as well as the photon and neutrino luminosities for each equation of state. We further examine the effects of the magnetic field, the choice of equation of state, and the underlying gravity theory framework on the cooling of neutron stars, particularly those of magnetized neutron stars or magnetars. Our results indicate that $f(R, T)$ gravity, particularly for the APR and SLy equations of state, exhibits improved agreement with the observed $T_s^{\infty}$ and photon luminosities than standard general relativity, regardless of magnetic-field strength. Moreover, it predicts the neutrino luminosities under both gravity models, all the chosen equations of state, and magnetic field configurations.
\end{abstract}

\maketitle

\section{Introduction}
\label{intro}

Magnetars and/or Neutron stars (NSs) are the most essential compact astrophysical objects for investigating the physics of ultra-dense matter under extreme gravitational fields and intense magnetic environments. The extreme densities of such astrophysical objects reaching and surpassing nuclear saturation render them unique laboratories for exploring matter under extreme conditions.~\cite {page2006cooling,potekhin2015neutron,wijnands2017cooling}. The thermal evolution of NSs determined by neutrino emission from the dense interior and photon emission from the stellar surface carries valuable information about their microphysical composition and the transport properties of the crust and core. Furthermore, this evolution influences observable quantities such as the red-shifted surface temperature \( T_s^{\infty} \) and their electromagnetic or neutrino luminosities~\cite{page2006cooling,potekhin2015neutron}. Canonical $1.4 \rm M_\odot$ neutron stars, in particular, serve as standard benchmarks for comparing theoretical predictions across different classes of equations of state (EoSs) and cooling scenarios. The cooling and emission properties of NSs also provide a sensitive tool for testing gravitational theories in regimes where gravity is strong and highly curved. In the conventional approach, cooling models rely on general relativity (GR), combined with the stellar equation of state, neutrino emissivities, and magnetic-field configurations that shape heat transport throughout the crust and core. However, several open questions in astrophysics and cosmology motivate the study of extended gravity theories. In view of the above, various extensions of GR have been proposed to address open problems in cosmology and high-energy physics, motivating efforts to examine their implications for the structure of compact stars. Among these frameworks, the $f(R,T)$ class of theories featuring a gravitational Lagrangian that depends on the Ricci scalar $R$ and the trace of the energy–momentum tensor $T$ introduces a non-minimal curvature-matter coupling that modifies the hydrostatic equilibrium equations~\cite{harko2011f}. This curvature–matter coupling modifies the Tolman-Oppenheimer-Volkoff (TOV) equations, alters the hydrostatic balance, and can influence mass–radius relations and red-shifted observables without violating spherical symmetry. Such modifications may leave detectable signatures in neutron-star cooling behavior, making thermal evolution a promising probe of deviations from GR.\\ 

Furthermore, magnetic fields play a crucial role in determining the thermal evolution of NSs. While typical pulsars possess surface magnetic fields of $10^{12}$–$10^{13}$ Gauss, and magnetars can exceed $10^{15}$ Gauss. Theoretical studies suggest that interior fields may reach values as high as $10^{18}$ Gauss. Such strong fields influence heat conduction, generate significant anisotropies in thermal transport, modify neutrino emission processes, and alter the relation between internal temperature and observable surface temperature through the magnetized heat blanketing envelope (HBEs)~\cite{chugunov2007thermal,potekhin2015neutron,beznogov2021heat}. Because both modified gravity and strong magnetic fields can independently reshape the internal structure of the star, their combined effect on thermal evolution is nontrivial and remains insufficiently explored in the literature. However, the EoS of dense matter and the envelope composition also significantly influence the observational appearance of cooling curves. Consequently, neutron-star thermal evolution is highly sensitive to changes in both gravitational dynamics and magnetic-field structure, making it an ideal setting to test the implications of $f(R, T)$ gravity.\\

Several previous studies have examined NS's cooling under different physical assumptions. In GR, magnetized cooling simulations using the NSCool code have provided robust and widely used predictions of red-shifted surface temperature $T_s^\infty$ and luminosities for various envelope compositions and nucleonic superfluid gaps~\cite{page2016nscool,beznogov2016cooling}. Modified gravity has also received attention. For instance, Nava-Callejas et al.~\cite{nava2023probing} studied NS cooling in $f(R)=R+\alpha R^2$ gravity, while scalar–tensor analogues have been shown to modify neutrino cooling channels and thresholds~\cite{dohi2021neutron}. In the context of compact stars of lower density, white-dwarf cooling has been analyzed within $f(R)$ and $f(R, T)$ frameworks, demonstrating gravity-dependent luminosity variations~\cite{bhattacharjee2024white,kalita2022cooling,kalita2023metric}. Despite these developments, neutron-star cooling, which incorporates both strong magnetic fields and curvature-matter coupling in the $f(R, T)$ framework, remains largely unexplored. It is essential to investigate such a scenario to explore how thermal emission can help in refining and constraining these gravity models. This inquiry could significantly enhance our understanding and lead to more accurate predictions.\\

Subsequently, we constructed a framework utilizing three widely recognized nucleonic equations of state; APR~\cite{akmal1998equation}, SLy~\cite{douchin2001unified}, and FPS~\cite{friedman1981hot}, which encompass a representative range of stiffness at super-nuclear densities. To simulate a strongly magnetized NS's interior, we incorporated a radially varying magnetic field profile with a central field strength of $B_c = 10^{18}$ Gauss. Throughout the framework, we employed consistent microphysical inputs, including neutrino emission from direct and modified Urca processes, Cooper-pair breaking and formation, and electron–electron bremsstrahlung in the crust~\cite{yakovlev2004neutron,haug2004elementary}. We also considered magnetic field-dependent thermal conductivity and opacity to depict magnetized transport through the heat-blanketing envelopes~\cite{chugunov2007thermal,potekhin2015neutron,beznogov2021heat}. Following these, the present study establishes a unified framework that integrates the effects of magnetic field and $f(R,T)$ corrections directly into the modified Tolman-Oppenheimer-Volkoff (M-TOV) equations. This approach facilitates a consistent derivation of mass, density, and pressure profiles, capturing the combined influences of curvature-matter coupling and strong magnetization. These equilibrium backgrounds serve as inputs for the NSCool simulation package~\cite{page2006cooling,page2016nscool}. This allows us to quantify the combined impact of magnetic fields and curvature-matter coupling on the red-shifted thermal observables, including the interior temperature $T_s^\infty(t)$, neutrino luminosity $L_\nu^\infty(t)$, and photon luminosity $L_\gamma^\infty(t)$ under both general relativity (GR) and $f(R, T)$ gravity.\\ 

The paper is organized and outlined as follows: Section~\ref{form} introduces the theoretical framework, starting with the spacetime metric and the derivation of the TOV equations within the context of general relativity (GR), and subsequently extending these concepts to the $f(R, T)$ gravity models. This section also provides a brief overview of the equations of state employed (APR, SLy, and FPS), as well as the radial magnetic-field profile. Furthermore, it discusses the implementation of these modifications in the NSCool code, which enables the computation of red-shifted observables such as temperature, neutrino luminosity, and photon luminosity. Section~\ref{result} details the computed cooling curves and luminosity evolutions for various combinations $(\lambda, B_c)$, comparing the results with general-relativistic baselines and selected observational data. Section~\ref{conc} delves into the physical interpretation of the findings, exploring the interaction between the magnetic field and curvature-matter coupling, as well as the degeneracy with respect to equations of state or envelope composition. Finally, we summarize the main findings and future prospects of the current work. 

\section{Formalism}
\label{form}
In order to connect the physical motivations with the quantitative analysis, this section establishes the mathematical framework that governs the structure and thermal evolution of magnetars. The formalism integrates the effects of modified gravity, dense-matter equations of state, and strong magnetic fields into a unified description of the pressure, density, and temperature distributions. By articulating the equilibrium equations and detailing how microphysical inputs are integrated into the cooling profiles, the subsections that follow provide the essential bridge between the conceptual picture outlined earlier and the methodological tools used to compute the observable thermal properties.

 \subsection{M-TOV equations for Non-rotating Magnetars}
\noindent We model a static, spherically symmetric, non-rotating star with the line element as:
\begin{equation}
ds^{2}=-e^{2\phi(r)}dt^{2}+\frac{dr^{2}}{1-2m(r)/r}+r^{2}d\Omega^{2}
\end{equation}

For an isotropic perfect fluid of density \(\rho\) and pressure \(P\), the TOV equations for GR are given as~\cite{tolman1939static,oppenheimer1939massive}:

\begin{equation}
\frac{dm}{dr} = 4\pi r^{2}\rho(r)
\label{eq1}
\end{equation}

\begin{equation}
\frac{d\phi}{dr} = \frac{G\left[m(r) + 4\pi r^{3}P/c^{2}\right]}{r\left(rc^{2} - 2Gm(r)\right)}
\label{eq2}
\end{equation}

\begin{equation}
\frac{dP}{dr} = -c^2\left(\rho(r) + P/c^2\right)\frac{d\phi}{dr}
\label{eq3}
\end{equation}

\noindent
These coupled equations are solved numerically with initial conditions; \(r=0\) with \(m(0)=0\), \(\rho(0)=\rho_c\), and proceeds to surface \(R\) where \(P(R)=0\), yielding \(M=m(R)\).\\

Now, taking the magnetic field into account leads to magnetic energy density \(\varepsilon_B=B^{2}/8\pi\). This magnetic energy density contributes to the mass source, along with a prescribed Lorentz-force profile \(\mathcal{L}(r)\) that aids in achieving hydrostatic balance, expressed as~\cite{dexheimer2017magnetic,chatterjee2015consistent}:

\begin{equation}
\mathcal{L}(r) = B_{c}^{2}\,[-3.8x + 8.1x^{3} - 1.6x^{5} - 2.3x^{7}]\times 10^{-41}
\qquad x = \frac{r}{\bar{r}}
\label{eq:Lorentz}
\end{equation}
\begin{equation}
B=B(r) = B_{c}\!\left[1 - 1.6x^{2} - x^{4} + 4.2x^{6} - 2.4x^{8}\right],
\label{eq:Bprofile}
\end{equation}
where, \(B_c\) is the central magnetic field and $\bar{r}$ is slightly larger than the NS's radius $R$.\\

The present study uses geometrized units ($G=c=1$, $\kappa=8\pi$) and adopts the modified linear gravity model $f(R, T)=R+2\lambda\kappa\, T$, which is an extension of $f(R)$ gravity with a trace $T$ of the energy-momentum tensor with Lagrangian matter $\mathcal{L}_m=-\rho$~\cite{nobleson2022comparison,harko2011f,sotiriou2010f,de2010f,pretel2021neutron,mahapatra2024neutron,moraes2016stellar,carvalho2017stellar}. As the electromagnetic tensor is traceless, $T=-\rho+3P$ involves matter only. Using the prescribed magnetic field $B(r)$ and Lorentz-force profile $\mathcal L(r)$ as considered in Refs.~\cite{rathod2025structural,yadav2024thermal,yadav2024x}, the modified TOV equations can be written as:

\begin{align}
\frac{dm}{dr} &= 4\pi r^2\!\left(\rho+\frac{B^2}{8\pi}\right) - \frac{r^2}{4}\;h(T)
\quad h(T)=2\lambda\kappa\,T \\
\frac{d\phi}{dr} &= \frac{m+4\pi r^3 P + \tfrac{\lambda\kappa}{2}\,r^3\!\left[T+2\!\left(\rho+\tfrac{B^2}{8\pi}+P\right)\right]}{r(r-2m)} \\
\frac{dP}{dr} &= -\!\left(\rho+P+\frac{B^2}{8\pi}\right)\!\left[\frac{d\phi}{dr}-\mathcal L(r)\right]
\end{align}\\

The above equations use the same initial conditions as chosen for Eqs.~\ref{eq1}, ~\ref{eq2} \& \ref{eq3}, along with additional consistency limits on $\lambda$ and $B$. At $(\lambda, B_c)\to\!(0,0)$, the solution of the modified TOV equations reduces to the GR solution. For only $\lambda\!\to\!0$, its provides a solution to the magnetized GR scenario and for $B_c\to 0$
solution corresponds to the non-magnetic $f(R,T)$ gravity.


\subsection{Equations of State}
\label{subsec:eos}

We have characterized neutron stars (NSs) using three different nucleonic equations of state (EoSs), such as APR~\cite{gusakov2005cooling, schneider2019akmal,akmal1998equation}, FPS~\cite{flowers1976neutrino}, and SLy~\cite{broderick2000equation}. The APR EoS is a variational many-body model that incorporates realistic two-nucleon and three-nucleon forces. It is relatively stiff, which implies that it produces high pressures at a relatively low density. In contrast, the FPS EoS provides a unified description of both the inner crust and the liquid core of neutron stars. According to this model, the transition from liquid core to crust occurs at a density of approximately \(\rho_{\mathrm{edge}} \simeq 1.6 \times 10^{14} \ \mathrm{g\,cm^{-3}}\). This model features first-order phase transitions, with density jumps typically below $1$\%, making the FPS EoS a representative of soft EoSs. The SLy EoS, on the other hand, is based on a Skyrme-type effective interaction that has been calibrated to neutron-rich matter. It provides a consistent description of the crust and core, yielding radii that lie between those of the APR and FPS EoSs. Together, these EoSs cover a useful range of stiffness to assess how \(f(R,T)\) coupling and strong magnetic fields affect the global structure and cooling of neutron stars.\\

In the \(f(R,T)\) framework, assuming near equipartition between magnetic and fluid pressures~\cite{chatterjee2015consistent,bocquet1995rotating}, Tables~\ref{tab:mass apr}, \ref{tab:mass fps}, and \ref{tab:mass sly} compile the maximum masses dependent on the equation of state (EoS) alongside the predicted radii.

\begin{table}[ht!]
\centering
\caption{Maximum mass and corresponding radius for different values of the modified gravity parameter \(\lambda\) for APR EoS, for \(B_c=0\) and \(B_c=10^{18}\,\mathrm{Gauss}\).}
\label{tab:mass apr}
\resizebox{0.4\textwidth}{!}{%
\begin{tabular}{|c|cc|cc|}
\hline
 & \multicolumn{2}{c|}{$B_c=0$} & \multicolumn{2}{c|}{$B_c=10^{18}\,\mathrm{Gauss}$} \\
\cline{2-5}
\(\lambda\) & \(\displaystyle M_{\rm max}\,(M_\odot)\) & \(R\;(\mathrm{km})\) 
            & \(\displaystyle M_{\rm max}\,(M_\odot)\) & \(R\;(\mathrm{km})\) \\
\hline
0                       & 2.20 & 10.07 & 2.17 & 9.85  \\
\(-\tfrac{1}{8\pi}\)   & 2.34 & 10.40 & 2.31 & 10.19 \\
\(-\tfrac{2}{8\pi}\)   & 2.51 & 10.86 & 2.49 & 10.64 \\
\(-\tfrac{3}{8\pi}\)   & 2.74 & 11.37 & 2.71 & 11.17 \\
\hline
\end{tabular}%
}
\end{table}

\begin{table}[ht!]
\centering
\caption{Maximum mass and corresponding radius for different values of the modified gravity parameter \(\lambda\) for FPS EoS, for \(B_c=0\) and \(B_c=10^{18}\,\mathrm{Gauss}\).}
\label{tab:mass fps}
\resizebox{0.4\textwidth}{!}{%
\begin{tabular}{|c|cc|cc|}
\hline
 & \multicolumn{2}{c|}{$B_c=0$} & \multicolumn{2}{c|}{$B_c=10^{18}\,\mathrm{Gauss}$} \\
\cline{2-5}
\(\lambda\) & \(\displaystyle M_{\rm max}\,(M_\odot)\) & \(R\;(\mathrm{km})\) 
            & \(\displaystyle M_{\rm max}\,(M_\odot)\) & \(R\;(\mathrm{km})\) \\
\hline
0                       & 2.05 & 10.24 & 2.03 & 9.98  \\
\(-\tfrac{1}{8\pi}\)   & 2.20 & 10.61 & 2.17 & 10.27 \\
\(-\tfrac{2}{8\pi}\)   & 2.38 & 11.03 & 2.35 & 10.75 \\
\(-\tfrac{3}{8\pi}\)   & 2.62 & 11.57 & 2.59 & 11.41 \\
\hline
\end{tabular}%
}
\end{table}

\begin{table}[ht!]
\centering
\caption{Maximum mass and corresponding radius for different values of the modified gravity parameter \(\lambda\) for SLy EoS, for \(B_c=0\) and \(B_c=10^{18}\,\mathrm{Gauss}\).}
\label{tab:mass sly}
\resizebox{0.4\textwidth}{!}{%
\begin{tabular}{|c|cc|cc|}
\hline
 & \multicolumn{2}{c|}{$B_c=0$} & \multicolumn{2}{c|}{$B_c=10^{18}\,\mathrm{Gauss}$} \\
\cline{2-5}
\(\lambda\) & \(\displaystyle M_{\rm max}\,(M_\odot)\) & \(R\;(\mathrm{km})\) 
            & \(\displaystyle M_{\rm max}\,(M_\odot)\) & \(R\;(\mathrm{km})\) \\
\hline
0                       & 2.05 & 10.03 & 2.02 & 9.76  \\
\(-\tfrac{1}{8\pi}\)   & 2.19 & 10.42 & 2.17 & 10.16 \\
\(-\tfrac{2}{8\pi}\)   & 2.37 & 10.88 & 2.35 & 10.57 \\
\(-\tfrac{3}{8\pi}\)   & 2.61 & 11.44 & 2.58 & 11.22 \\
\hline
\end{tabular}%
}
\end{table}

\subsection{Cooling of Neutron Stars}
\label{nsc}
The primary mechanism responsible for the cooling of neutron stars (NS) during their early stages and high-temperature phase is the emission of neutrinos and photons~\cite{wijnands2017cooling,potekhin2015neutron}. Here, the cooling is obtained using the NSCool code, which is a computational tool designed to model the thermal evolution of neutron stars and to conduct cooling simulations of compact stars~\cite{page2006cooling,page2016nscool,brown1988strangeness}. NSCool code incorporates equations for energy balance and heat transport, considering various cooling mechanisms such as neutrino emission, photon radiation, and heat conduction~\cite{buschmann2022upper,iwamoto1995neutrino,prakash1994rapid} through the heat-bearing envelope.\\

The heat transport under the isothermal limit is defined as;

\begin{equation}
C_v \frac{dT_b^\infty}{dt} = -L_\nu^\infty(T_b^{\infty}) - L_\gamma^\infty(T_s) + H
\end{equation}

\noindent The quantities \( C_v \), \( L_\nu^\infty \), and \( L_\gamma^\infty \) refer to the specific heat, neutrino luminosity, and surface photon luminosity, respectively. The symbol \( H \) denotes all potential contributions from heating processes occurring outside the neutron star (NS)~\cite{beznogov2023standard}. The temperature at the surface is represented as \( T_s \), while \( T_b^\infty \) indicates the temperature at any internal location. The photon luminosity can be expressed as \( L_\gamma^\infty = 4 \pi \sigma R^2 (T_s^\infty)^4 \), where \(\sigma\) is the Stefan-Boltzmann constant, and \( R \) is the radius of the star.\\ 

Since the external observer is situated at infinity, they measure these quantities on a redshifted scale, which is why the superscript \( \infty \) is used here. The relation \( T_s^\infty / T_s \sim 0.7 \) is generally applied. The relationship between the red-shifted temperature (as measured on Earth) and the surface temperature is expressed as 
\( T_s^\infty = T_s \sqrt{1 - \frac{2GM}{c^2R}} \). In this context, we assume \( H = 0 \). 
We have considered that the core matter inside the neutron star exists in a superfluid state, provided that the internal temperature of the neutron star is equal to or lower than the superfluid critical temperature.\\

In a scenario where the thermal relaxation timescale of the envelope is significantly shorter than the stellar evolution timescale and neutrino emission from the envelope is negligible, we can use ordinary differential equations to describe heat transport and hydrostatic equilibrium within the envelope. Once the microphysical inputs are established, we can solve these equations to determine a corresponding surface temperature and effective temperature of the envelope, denoted as \(T_s \)\ and \(T_e\) ($T_s \approx T_e$), respectively, for each internal temperature represented as \(T_b \equiv T(\rho_b)\). This relationship is commonly referred to as the \(T_b - T_s\) or \(T_b - T_e\) relation. The consistent observation that 
\(T_e\) increases monotonically with \(T_b\) is in line with findings reported in the Refs.~\cite{buschmann2022upper,gudmundsson1983structure,nomoto1987cooling}

\subsection{Radiative Opacity and Thermal Conductivity of HBEs}
The HBEs in NSs act as thermal insulators, regulating the transfer of heat from the stellar interior to the surface. These envelopes are composed of either pure iron or layered structures that contain hydrogen, helium, carbon, and iron. The choice of composition influences the thermal properties of the NSs, particularly their radiative opacity and heat conductivity, which in turn affect their cooling behavior.

\subsubsection{Effects of Magnetic Fields on Radiative Opacity}
The radiative opacity  for a fully ionized, non-relativistic, and non-degenerate plasma can be expressed as~\cite{beznogov2021heat}:

\begin{equation}
K_r (0) = 75\; g_{\text{eff}}\; \frac{Z^3}{A^2} \rho \left( \frac{10^6 K}{T} \right)^{3.5}
\label{krmer}
\end{equation}

where \( T \) is the temperature in Kelvin, \( g_{\text{eff}} \sim 1 \) is the effective Gaunt factor, \( Z \) and \( A \) are the atomic number and atomic mass of the ions, respectively. Equation~\ref{krmer} implies that opacity increases with higher atomic number elements, suggesting that an iron-dominated envelope exhibits greater opacity. In contrast, an envelope containing lighter elements, such as hydrogen and helium, results in lower opacity, leading to more efficient heat transport. The effects of the high magnetic fields in the outer layers of the NSs affect the effective surface temperature and the heat flow near the stellar surface~\cite{beznogov2021heat}.

\begin{equation}
K_r(B) = 2.2 g_{\text{eff}} Z^3 \rho A^{-2} \left( \frac{10^6 K}{T} \right)^{1.5} \left( \frac{10^{12}\, \mbox{Gauss}}{B} \right)^2
\end{equation}

\begin{equation}
K_r(B) = K_r(0) \left(\frac{T_6}{B_{12}}\right)^2
\end{equation}

The heat flux through the envelope is governed by Fourier’s law:
\begin{equation}
F_r = -\kappa^c \frac{dT}{dz}
\end{equation}

where, \( F_r \) is the radial heat flux, \( \kappa^c \) is the thermal conductivity, and \( z \) represents the depth from the surface. The relationship between the internal temperature (\( T_b \)) and the effective surface temperature (\( T_s \)) is often approximated by:

\begin{equation}
T_s^4 = g_s f(T_b)
\end{equation}
where, \( g_s \) is the surface gravity and \( f(T_b) \) is a function that depends on the composition of the envelope.

\subsubsection{Effects of Magnetic Fields on Thermal Conductivity}
In the presence of a strong magnetic field, the thermal conductivity of the heat-blanketing layer becomes anisotropic. The parallel and transverse components of the electron thermal conductivity are  expressed~\cite{potekhin2001thermal,chugunov2007thermal} as:

\begin{equation}
\kappa^c_{\parallel} = \frac{\pi^2 k_B^2 T n_e \tau_e}{3m^*_e}
\end{equation}

\begin{equation}
\kappa^c_{\perp} = \frac{\kappa^c_{\parallel}}{1 + (\omega^* \tau_e)^2},
\end{equation}

where, \( \tau_e \) is the relaxation time associated with effective electronic thermal conduction, \( n_e \) is the density of the electron number, \( k_B \) is Boltzmann's constant, and \( m_e \) is the electron mass. The quantity \( \omega^* \) is given by:

\begin{equation}
\omega^* = \frac{\omega_c}{\gamma_r}, \quad \omega_c = \frac{eB}{m_e c}, \quad \gamma_r = \sqrt{1 + x_r^2}, \quad x_r = \frac{p_F}{m_e c}
\end{equation}

The product \( \omega^* \tau_e \) can be approximated as:
\begin{equation}
\omega^* \tau_e = 1760 \left( \frac{B_{12}}{\gamma_r} \right) \left( \frac{\tau_e}{10^{-16}s} \right),
\end{equation}

where, \( B_{12} = B/10^{12} \) Gauss, and \( \theta_B \) is the angle between the magnetic field direction and the normal to the stellar surface.

The presence of a magnetic field alters heat transport by suppressing conduction in the direction perpendicular to the field, while allowing efficient heat flow along the field lines. This anisotropic heat transfer impacts neutron star cooling and surface temperature distribution, making magnetic effects crucial in modeling the thermal evolution of neutron stars.

\subsection{Neutrino emission inside NS core}
\label{aer}
\subsubsection{Cooper Pair Breaking and Formation process (PBF)}
\label{cooper}
The pair-breaking and formation (PBF) process operates within the superfluid core of neutron stars~\cite{page2009neutrino}. As the stellar core cools below the critical temperature $T_c$, the baryonic dense matter undergoes a phase transition into a superfluid state. This transition is characterized by the formation of Cooper pairs among baryons, analogous to the pairing mechanism in conventional superconductors. In the superfluid state, thermal excitations within the star can break these Cooper pairs. The subsequent re-formation of Cooper pairs leads to the emission of neutrino–antineutrino pairs, providing an efficient channel for energy loss~\cite{yakovlev2001neutrino}.

The PBF process serves as a crucial cooling mechanism, especially during the evolution of the NSs. This process becomes dominant when the temperature decreases sufficiently for other neutrino-emission processes to be significantly suppressed. The PBF mechanism can be described as:

\begin{align} 
n+n\to \left [ nn \right ] + \nu +\overline{\nu}\nonumber\\
p+p\to \left [ pp \right ] + \nu +\overline{\nu}\nonumber
\end{align}
 
When the temperature falls below the critical temperature \( T_c = 10^9 \, \text{K} \) of neutron (and proton) superfluid, axial-vector currents drive the significant cooling of neutron stars (NSs) with superfluid inside the core by the neutron/proton-wave coupled of neutrinos in a neutron/proton \(^1S_0 \)-wave coupled superfluid is determined in ref.~\cite{leinson2000neutrino} and is given by:

\begin{equation}
\epsilon_\nu^s = \frac{5 G_F^2}{14 \pi^3} \nu_N(0) v_F(N)^2 T^7 I_\nu^s
\end{equation}

Here, the integral \( I_\nu^s \) is:
\begin{equation}
I_\nu^s = z_N^7 \left( \int_1^{\infty} \frac{y^5}{\sqrt{y^2 - 1}} \left[ f_F(z_N y) \right]^2 dy \right)
\end{equation}
where, \( \epsilon_\nu^s \) is the neutrino emissivity, \( G_F \) is Fermi's coupling constant (\( G_F = 1.166 \times 10^{-5} \, \text{GeV}^{-2} \)), \( z = \Delta(T) / T \) with \( \Delta(T) = 3.06 T_c \sqrt{1 - T/T_c} \). Here \( T_c \) is the critical neutron/proton superfluid temperature.

\subsubsection{Direct Urca (Durca) Processes}
When the proton fraction ($Y_p$) in neutron star matter exceeds a critical threshold (typically $Y_p \gtrsim 0.11$), the energy and momentum conservation requirements are satisfied and the Direct Urca neutrino emission process is allowed kinematically. This mechanism significantly enhances neutrino emissivity, thereby accelerating the cooling of the NSs.

\noindent The direct Urca process fundamentally consists of two beta-decay reactions involving neutrons ($n$), protons ($p$), electrons ($e^-$), electron neutrinos ($\nu_e$), and electron antineutrinos ($\bar{\nu}_e$):
\begin{align*}
    n \rightarrow p + e^- + \bar{\nu}_e, \quad
    p + e^- \rightarrow n + \nu_e
\end{align*}

\noindent In the first reaction, a neutron decays into a proton, an electron, and an electron antineutrino. In contrast, the second reaction involves the electrons captured by a proton to produce a neutron and an electron neutrino. The continuous occurrence of these reactions results in significant neutrino emission, which effectively transports energy away from the star, thereby contributing to its cooling.

\noindent According to Yakovlev et al.~\cite{yakovlev1999cooling}, the neutrino emissivity $Q_{\text{Durca}}$ due to the direct Urca process can be expressed as:

\begin{equation}
    Q_{\text{Durca}} = 4.24 \times 10^{27} \, r_n \, r_p \left(\frac{T}{10^9\,\text{K}}\right)^6 \left(\frac{Y_e \rho}{0.16}\right)^{1/3}\,
    \text{erg\,cm}^{-3}\text{s}^{-1}
\end{equation}
where, \textbf{$r_n$, $r_p$} are dimensionless effective mass correction factors for neutrons and protons, respectively. These factors account for the modification of nucleon masses by nuclear interactions in dense neutron star matter. These are given by:
\begin{equation}
r_n = \frac{m_n^*}{m_n}, \quad r_p = \frac{m_p^*}{m_p}
\end{equation}
where, $m_n^*$ and $m_p^*$ represent effective neutron and proton masses within neutron star matter, and $m_n$, $m_p$ are their free-space (vacuum) masses. \textbf{$\left(\frac{Y_e \rho}{0.16}\right)^{1/3}$} explains the dependence on the electron fraction ($Y_e$) and the mass density $\rho$.

\noindent In neutron stars with magnetic fields stronger than the quantum critical field ($B \gtrsim 4.4 \times 10^{13}\,\text{Gauss}$), charged particle motions become quantized into discrete energy states, known as Landau levels. For ultra-strong fields ($B \gtrsim 10^{16}\,\text{Gauss}$), this quantization significantly modifies particle distributions and reaction rates, thus affecting the direct Urca neutrino emission process. Depending on the magnetic field strength and orientation, neutrino emissivity can be notably enhanced or suppressed. Accurate treatment of these magnetic effects requires specialized theoretical studies, as described by Baiko and Yakovlev in ref.~\cite{baiko1998direct}.

\subsubsection{Modified Urca (Murca) Processes}

In typical neutron stars with masses $M \sim 1.4\,M_\odot$, the proton fraction $Y_p$ in the core usually remains below the critical value required for the direct Urca process. In this regime, energy and momentum conservation cannot be satisfied by a simple beta decay, and the dominant neutrino-emission mechanism is the modified Urca (Murca) process~\cite{friman1979neutrino,yakovlev2001neutrino,yakovlev2004neutron}. An additional ``spectator'' nucleon participates in the reaction, carrying away the excess momentum and thereby allowing beta transitions to proceed even when direct Urca is kinematically forbidden. As a result, Murca processes operate over a broad density range and govern the standard neutrino cooling of $1.4\,M_\odot$ neutron stars.\\

\noindent The nucleonic modified Urca reactions consist of two branches involving neutrons ($n$), protons ($p$), electrons ($e^-$), electron neutrinos ($\nu_e$), and electron antineutrinos ($\bar{\nu}_e$) described as follows:
\begin{align*}
    &\text{Neutron branch:} \\
    &n + n \rightarrow n + p + e^- + \bar{\nu}_e\\
    &n + p + e^- \rightarrow n + n + \nu_e \\
    &\text{Proton branch:} \\
    &p + n \rightarrow p + p + e^- + \bar{\nu}_e \\
    &p + p + e^- \rightarrow p + n + \nu_e
\end{align*}

 In each cycle, a beta decay (or its inverse) converts a neutron into a proton (or vise-versa), while the spectator nucleon ensures momentum conservation in the strongly degenerate Fermi sea. The repeated operation of these reactions produces a continuous outflow of neutrinos, which removes thermal energy from the stellar interior, albeit at a lower rate than in the direct Urca channel.\\

Following the standard parameterization of Friman and Maxwell~\cite{friman1979neutrino}, Yakovlev et al.~\cite{yakovlev2001neutrino} implemented in the NSCool code~\cite{page2016nscool}, the neutrino emissivity due to the neutron-branch modified Urca process in non-superfluid matter can be written in the convenient form:
\begin{eqnarray}
    Q_{\text{Murca}}^{(n)} \simeq 2 \times 10^{21}\,
    R_{\text{Murca}}(\rho,Y_p,m_n^*,m_p^*) \\\nonumber
    \left(\frac{T}{10^9\,\text{K}}\right)^8
    \,\text{erg\,cm}^{-3}\,\text{s}^{-1},
\end{eqnarray}
where $T$ is the local temperature and \textbf{$R_{\text{Murca}}$} is a dimensionless factor of order unity that encapsulates the dependence on baryon density $\rho$, composition (proton and lepton fractions), and effective nucleon masses $m_n^*$ and $m_p^*$. These are utilized in the NSCool code. The proton branch contributes a comparable amount and is treated in an analogous way. In practice, NSCool sums over both branches to obtain the total emissivity as a result of modified Urca processes.\\

 The strong temperature dependence ($Q_{\text{Murca}} \propto T^8$) implies that modified Urca cooling is much slower than the direct Urca channel ($\propto T^6$ with a significantly larger prefactor), but it still dominates over the other neutrino luminosities in standard-cooling of $1.4\,\rm M_\odot$ stars, where direct Urca is kinematically forbidden~\cite{yakovlev2004neutron,beznogov2023standard}. 


\subsection{Neutrino emission inside NS crust}
\label{nnbrem}
\subsubsection{e-e Bremsstrahlung process}
\label{e-e}
\subsubsection*{Neutrino emission rate}
The e-e bremsstrahlung mechanism contributes significantly to cooling in the crust of a neutron star, especially in the early phases of the star's life cycle. The electron gas becomes relativistic in the densest regions of the crust, where electron degeneracy occurs, allowing electrons to occupy states up to the Fermi energy. Electron-electron scattering results from the electromagnetic interaction between the electrons in this environment. The electrons emit virtual photons during these scattering events, and the extra energy is released as a neutrino-antineutrino pair.
This weak interaction-driven process enables neutrinos to carry energy away from the star, resulting in its cooling. The e-e bremsstrahlung contributes sizably at higher temperatures and electron degeneracies. In neutron stars, thermal relaxation between the crust and core is greatly accelerated by the electron-electron bremsstrahlung process, particularly in the inner crust where densities are highest. Neutron stars typically have high temperatures $T = 10^8\,\mathrm{K}$ during the early cooling phase. For young neutron stars and accreting neutron stars, where accretion causes the crust to grow hot. Nevertheless, this cooling process is of paramount importance and is described in Ref.~\cite {ofengeim2014neutrino}. 

The neutrino emission rate for the e-e bremsstrahlung process is given by~\cite{kaminker1999neutrino}:
\begin{equation}
Q_{ee} = \frac{\pi^4 \xi_1 G_F^2 e^4 C_{e+}^2}{378 \hbar^9 c^{10} y_s p_{F_e}^2} n_e (k_B T)^8
\end{equation}

\begin{equation}
\approx \frac{0.69 \times 10^{14}}{y_s} \left(\frac{n_e}{n_0} \right)^{1/3} T_9^8 \quad \text{erg cm}^{-3} \text{s}^{-1}
\end{equation}

\section{Results and Discussions}
\label{result}

\begin{figure*}[t]
  \centering

  \begin{subfigure}{0.50\textwidth}
    \centering
    \includegraphics[width=\linewidth]{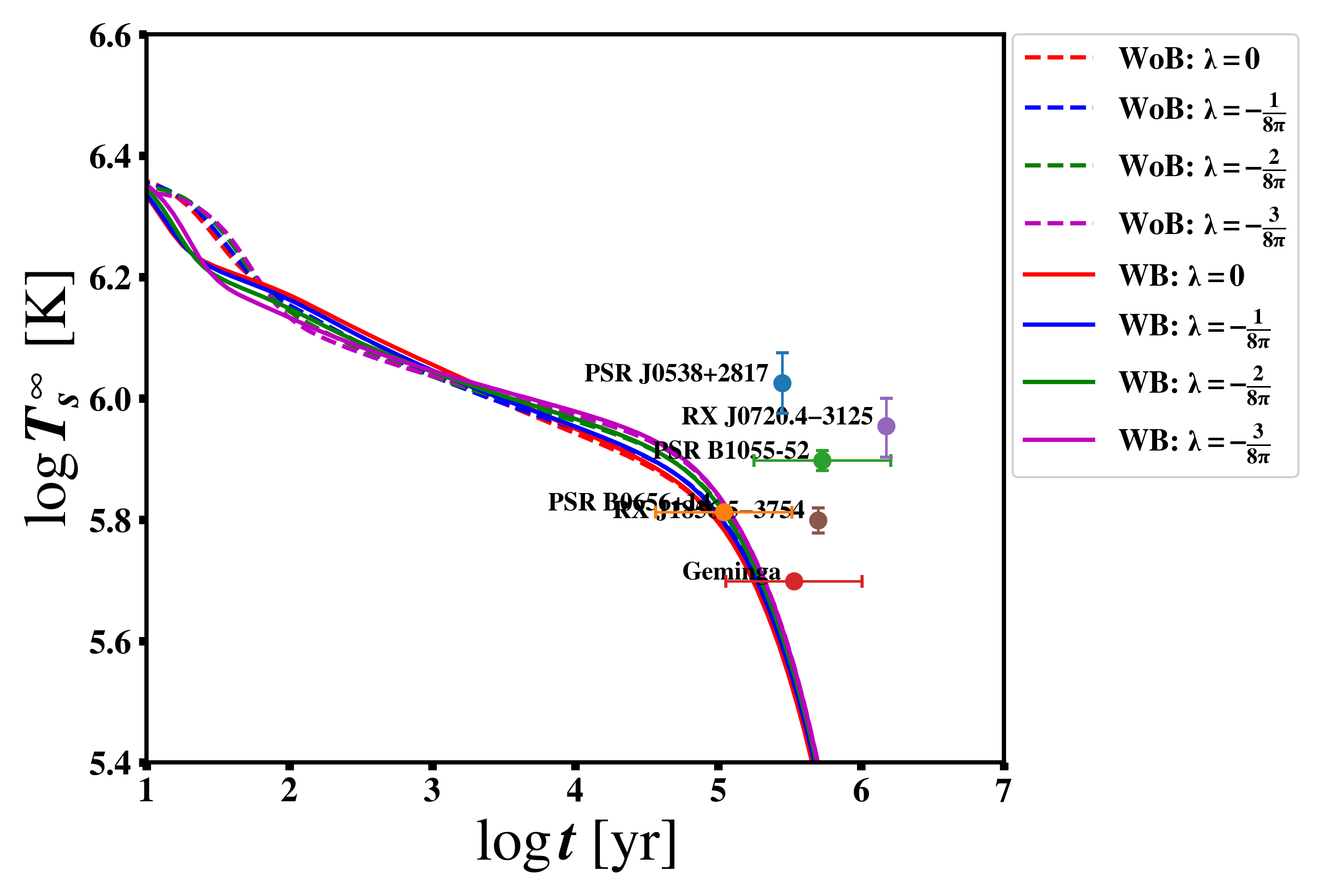}
    \caption{}
  \end{subfigure}

  \medskip

  \begin{subfigure}{0.50\textwidth}
    \centering
    \includegraphics[width=\linewidth]{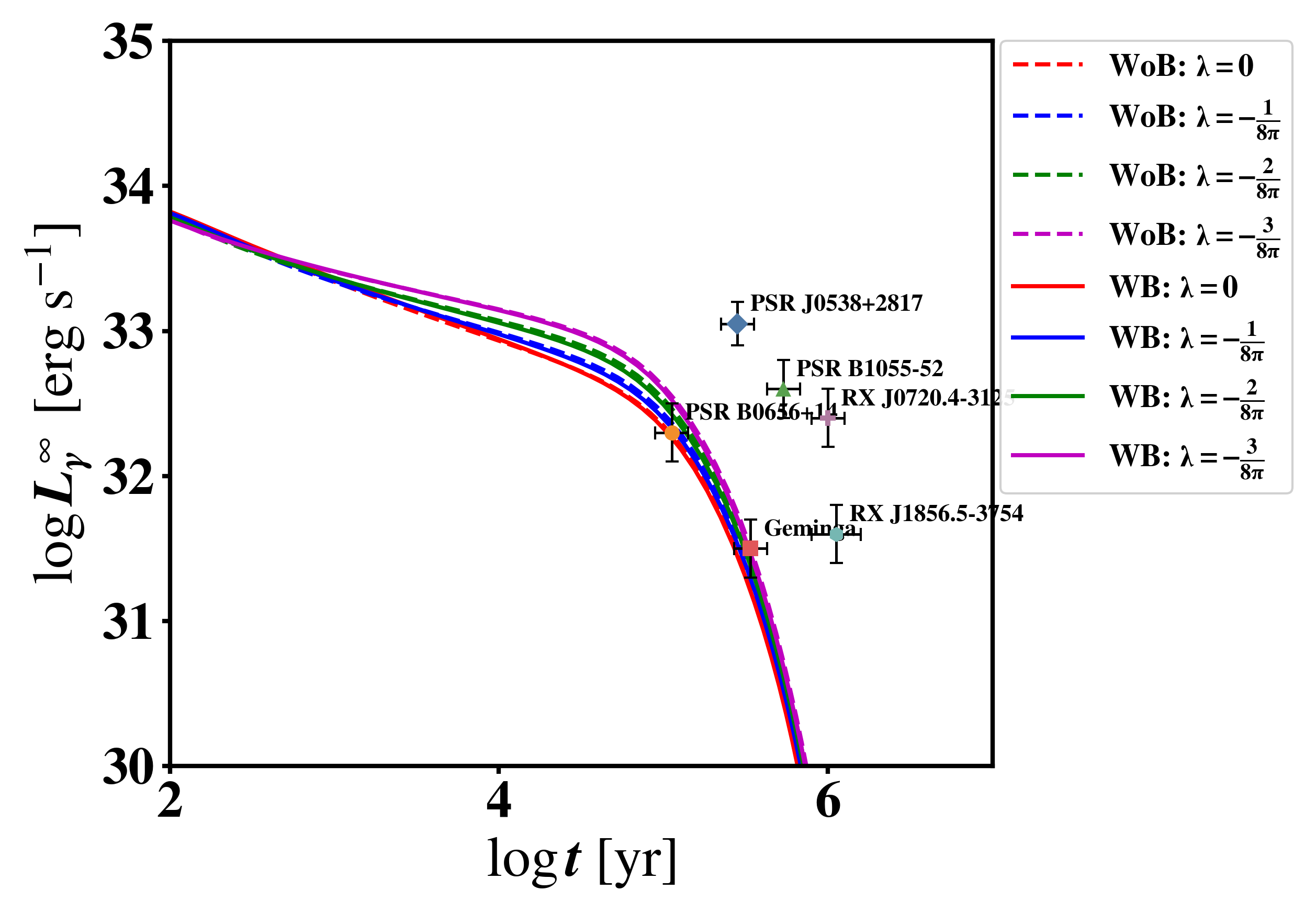}
    \caption{}
  \end{subfigure}\hfill
  \begin{subfigure}{0.50\textwidth}
    \centering
    \includegraphics[width=\linewidth]{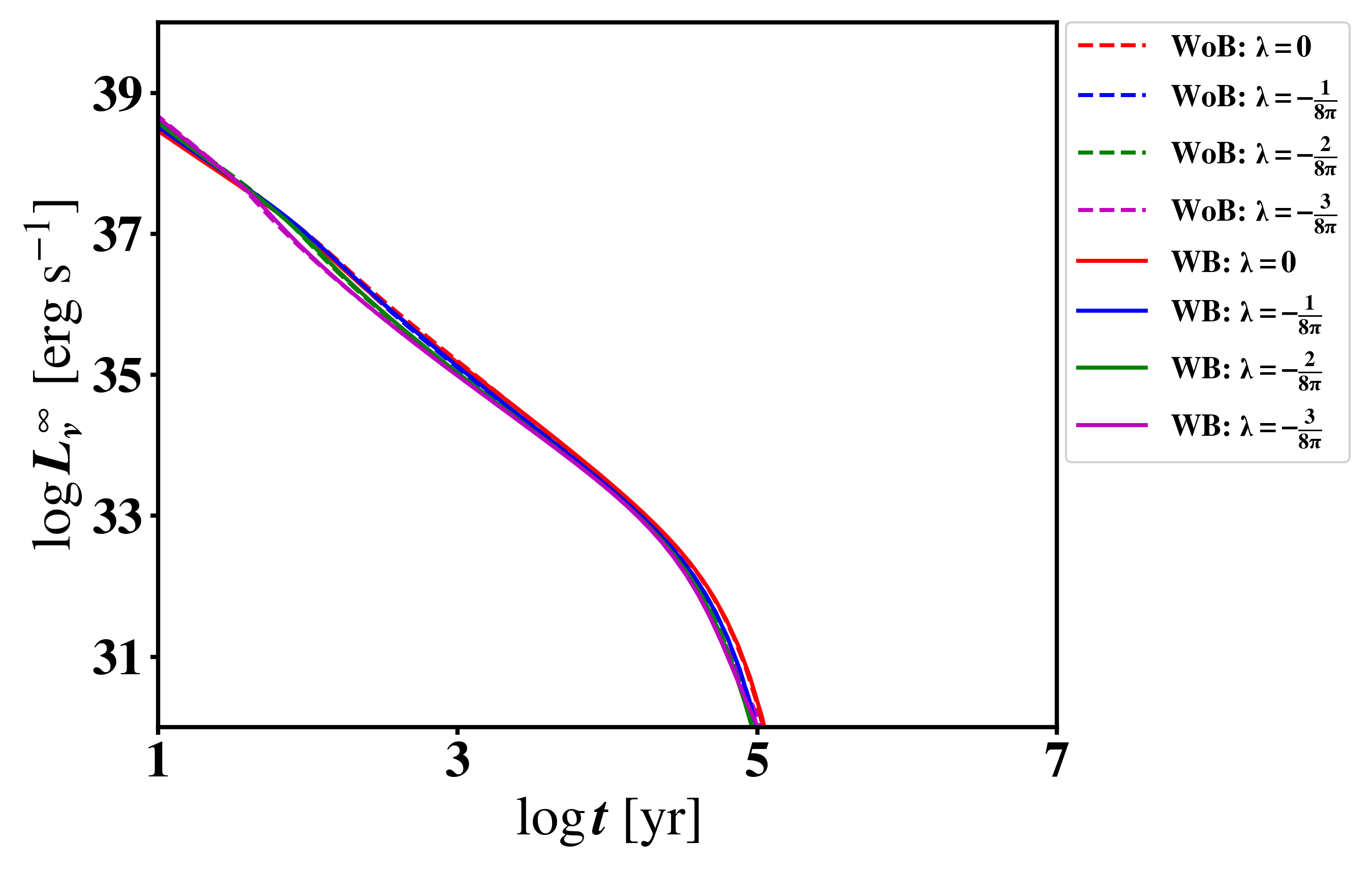}
    \caption{}
  \end{subfigure}

\caption{The variation of (a) red-shifted temperature ($T_s^\infty$), (b) red-shifted luminosity of photon ($L_\gamma^\infty$) and (c) red-shifted luminosity of neutrino ($L_\nu^\infty$) versus time for APR EoS with and without magnetic field. Observed data have also been shown for comparison.}
  \label{fig:apr}
\end{figure*}

\begin{figure*}[t]
  \centering

  \begin{subfigure}{0.50\textwidth}
    \centering
    \includegraphics[width=\linewidth]{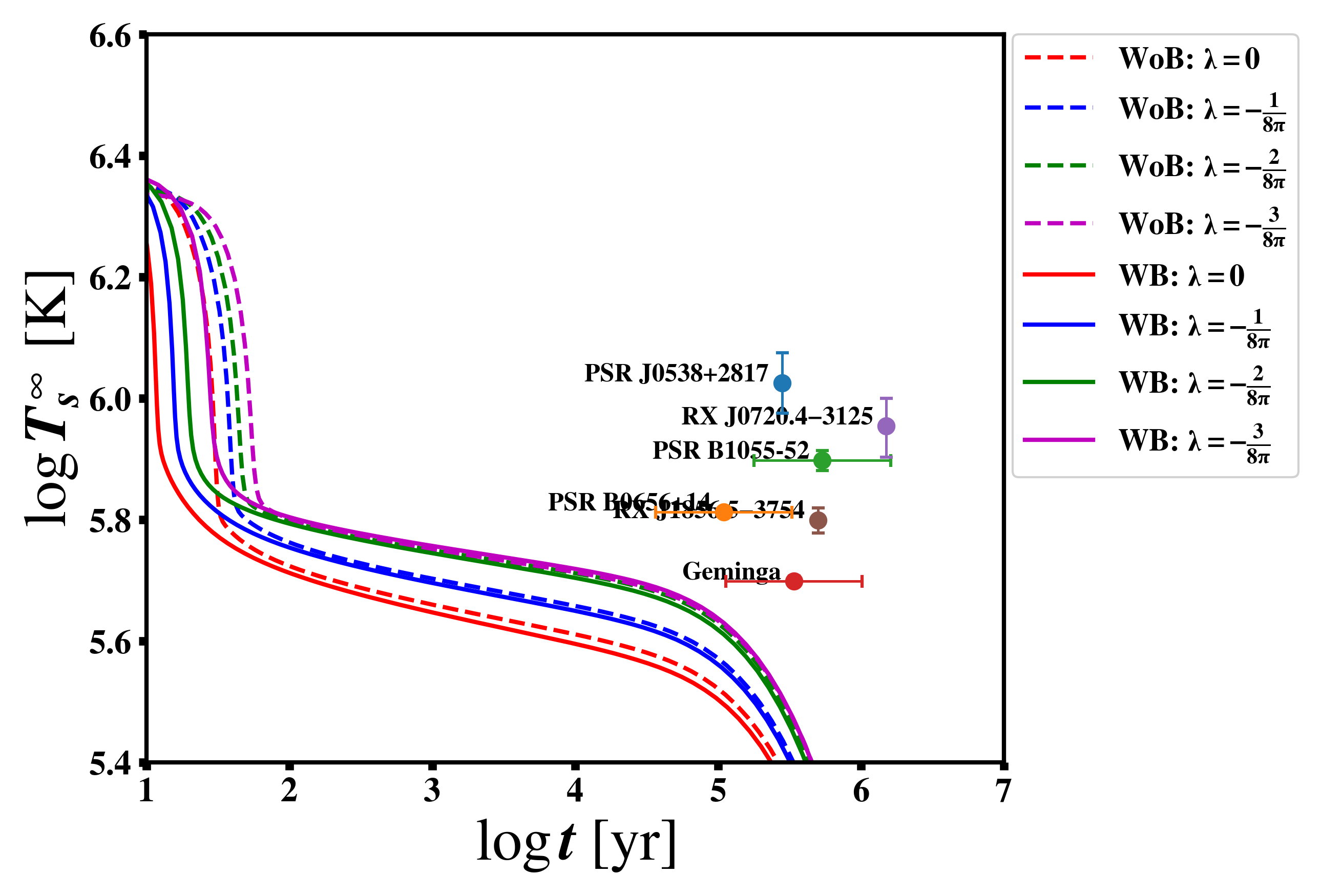}
    \caption{}
  \end{subfigure}

  \medskip

  \begin{subfigure}{0.50\textwidth}
    \centering
    \includegraphics[width=\linewidth]{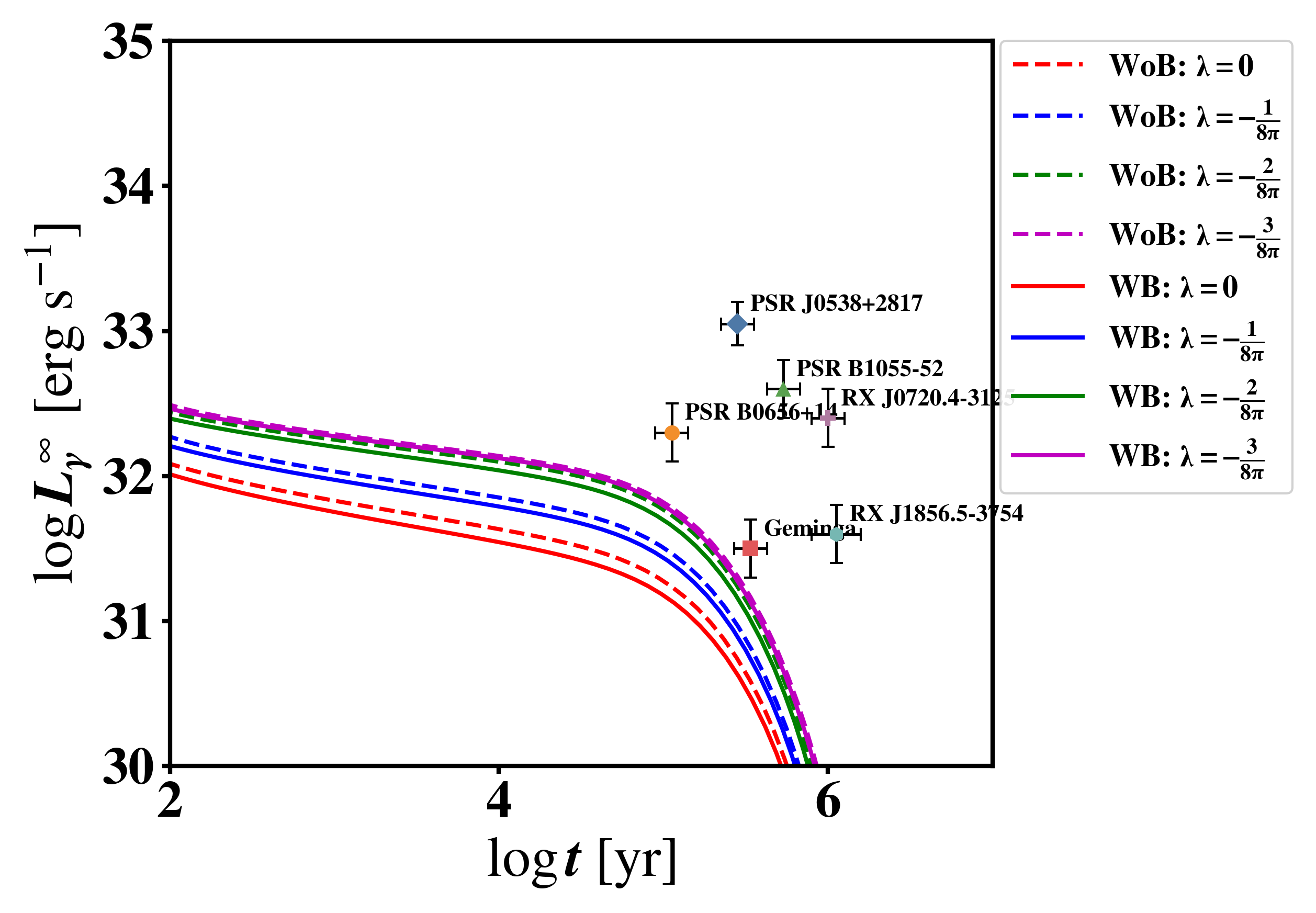}
    \caption{}
  \end{subfigure}\hfill
  \begin{subfigure}{0.50\textwidth}
    \centering
    \includegraphics[width=\linewidth]{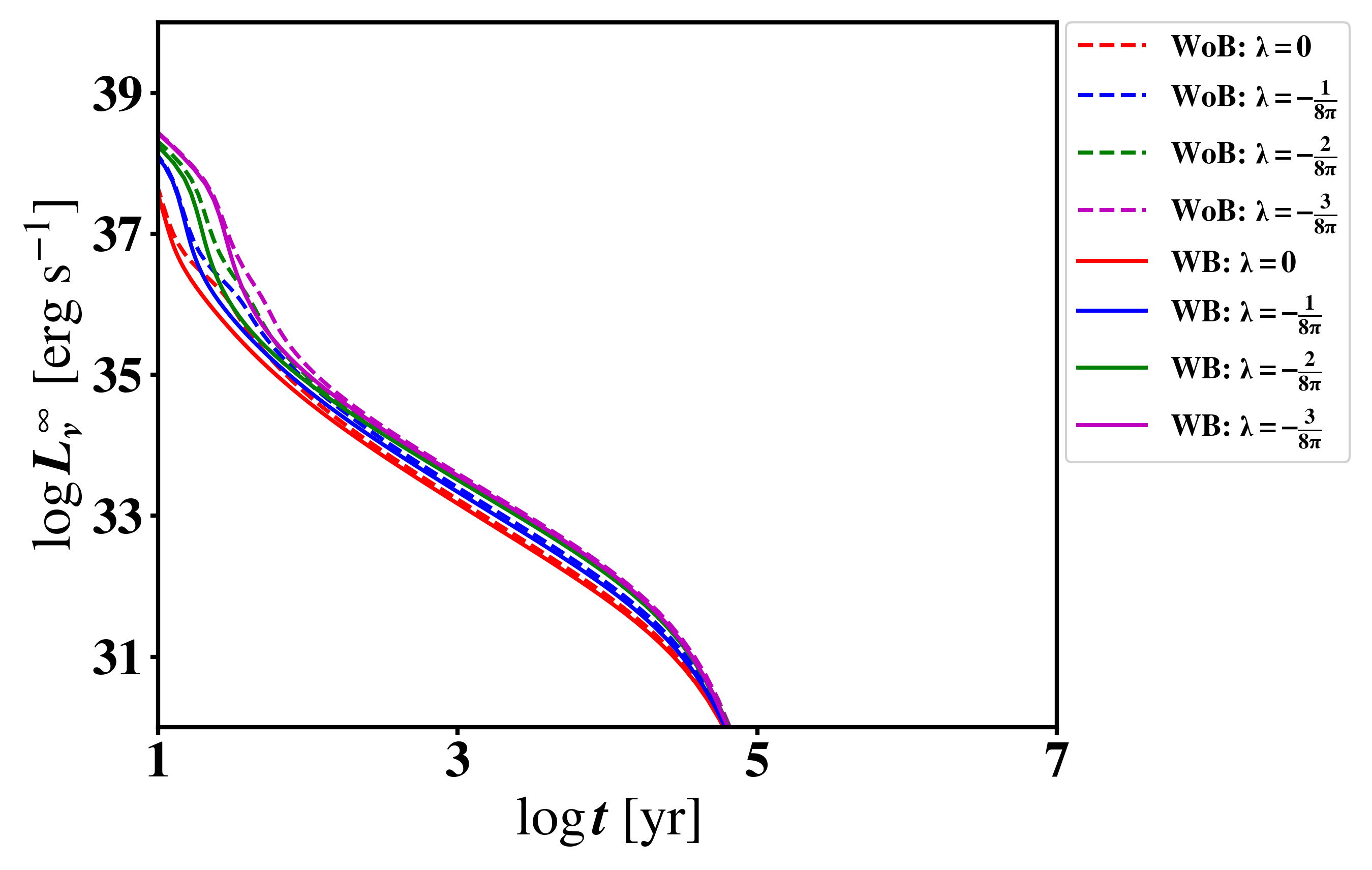}
    \caption{}
  \end{subfigure}

  \caption{The variation of (a) red-shifted temperature ($T_s^\infty$), (b) red-shifted luminosity of photon ($L_\gamma^\infty$) and (c) red-shifted luminosity of neutrino ($L_\nu^\infty$) versus time for FPS EoS with and without magnetic field. Observed data have also been shown for comparison.}
  \label{fig:fps}
\end{figure*}
\begin{figure*}[t]
  \centering

  \begin{subfigure}{0.50\textwidth}
    \centering
    \includegraphics[width=\linewidth]{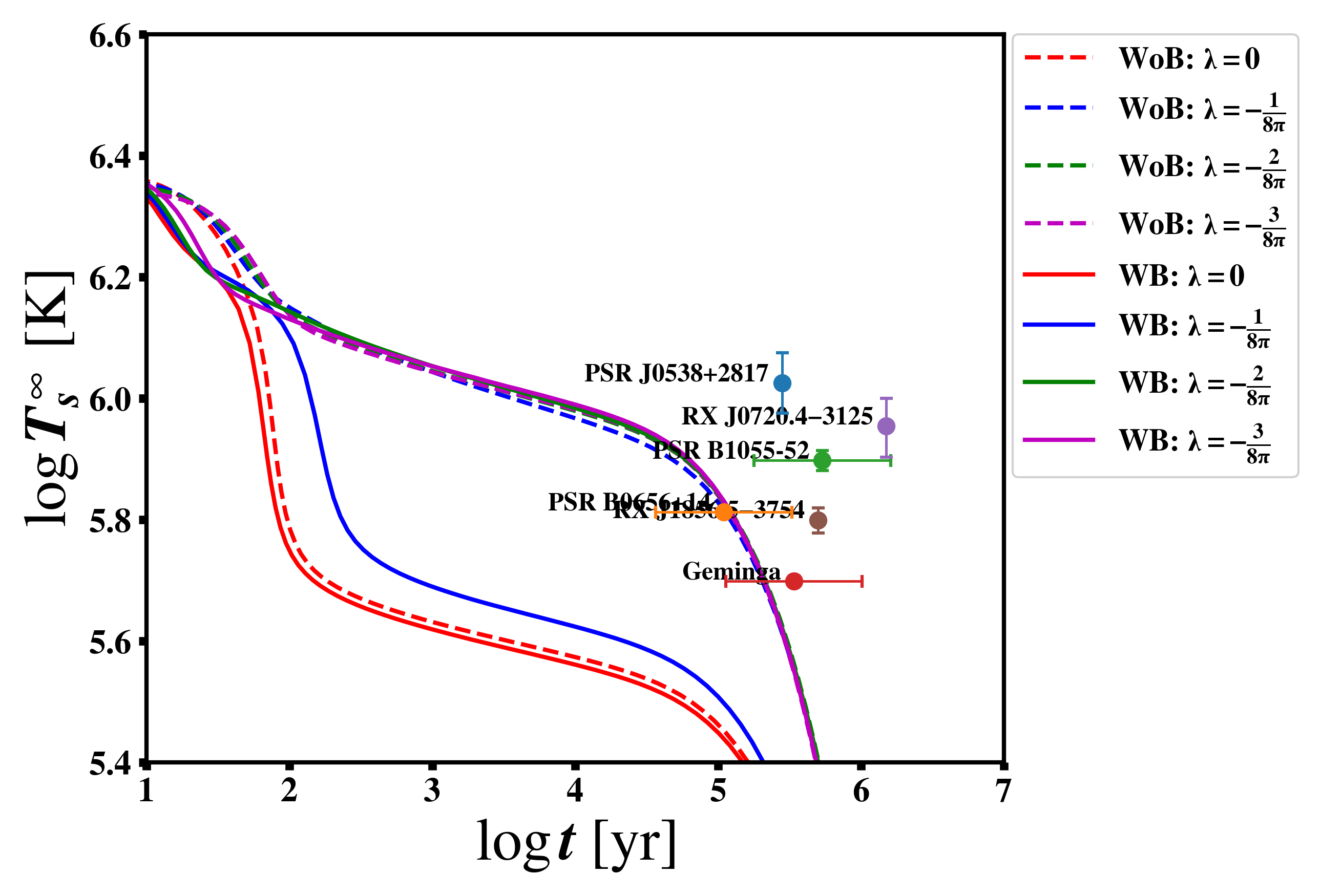}
    \caption{}
  \end{subfigure}

  \medskip

  \begin{subfigure}{0.50\textwidth}
    \centering
    \includegraphics[width=\linewidth]{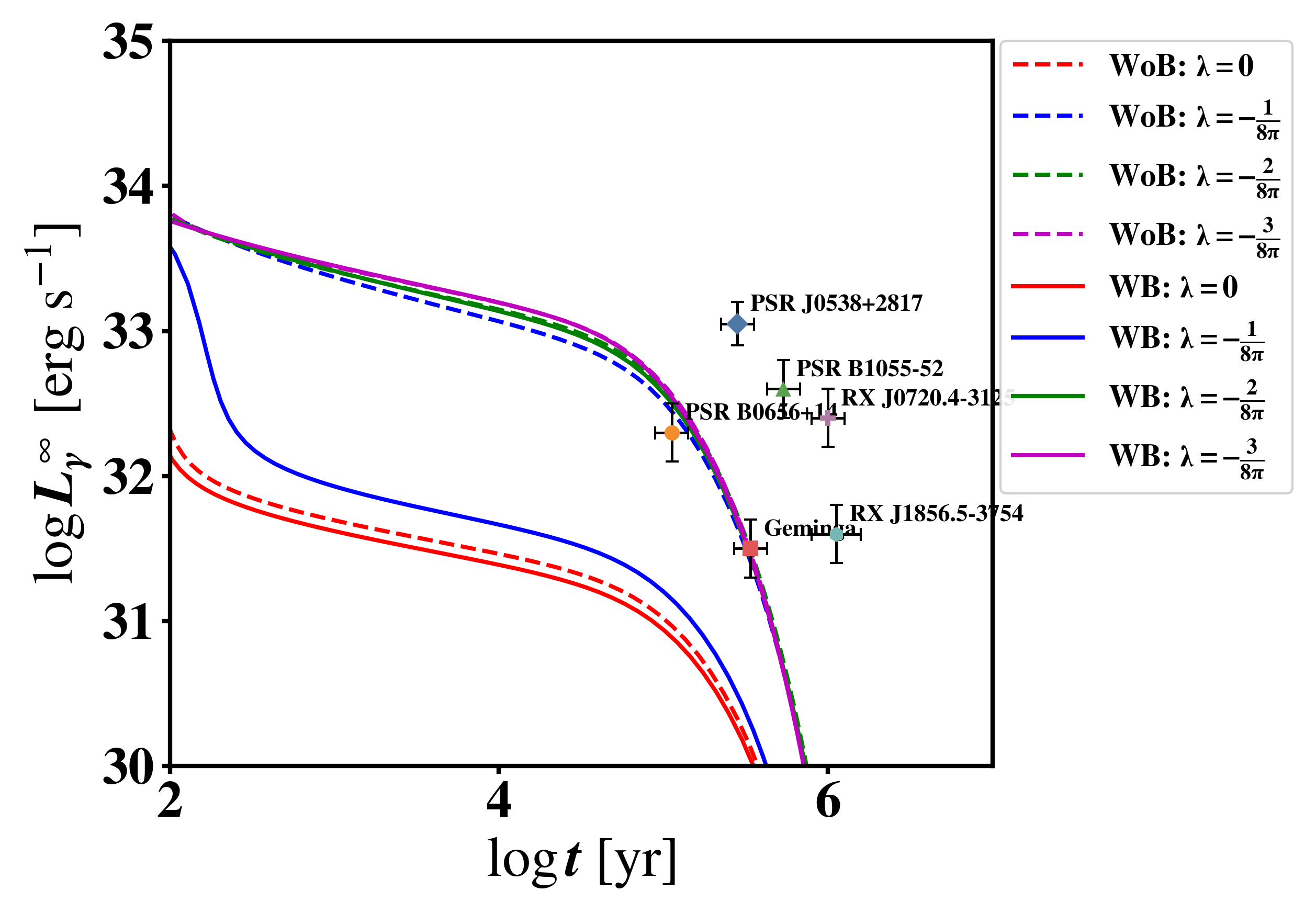}
    \caption{}
  \end{subfigure}\hfill
  \begin{subfigure}{0.50\textwidth}
    \centering
    \includegraphics[width=\linewidth]{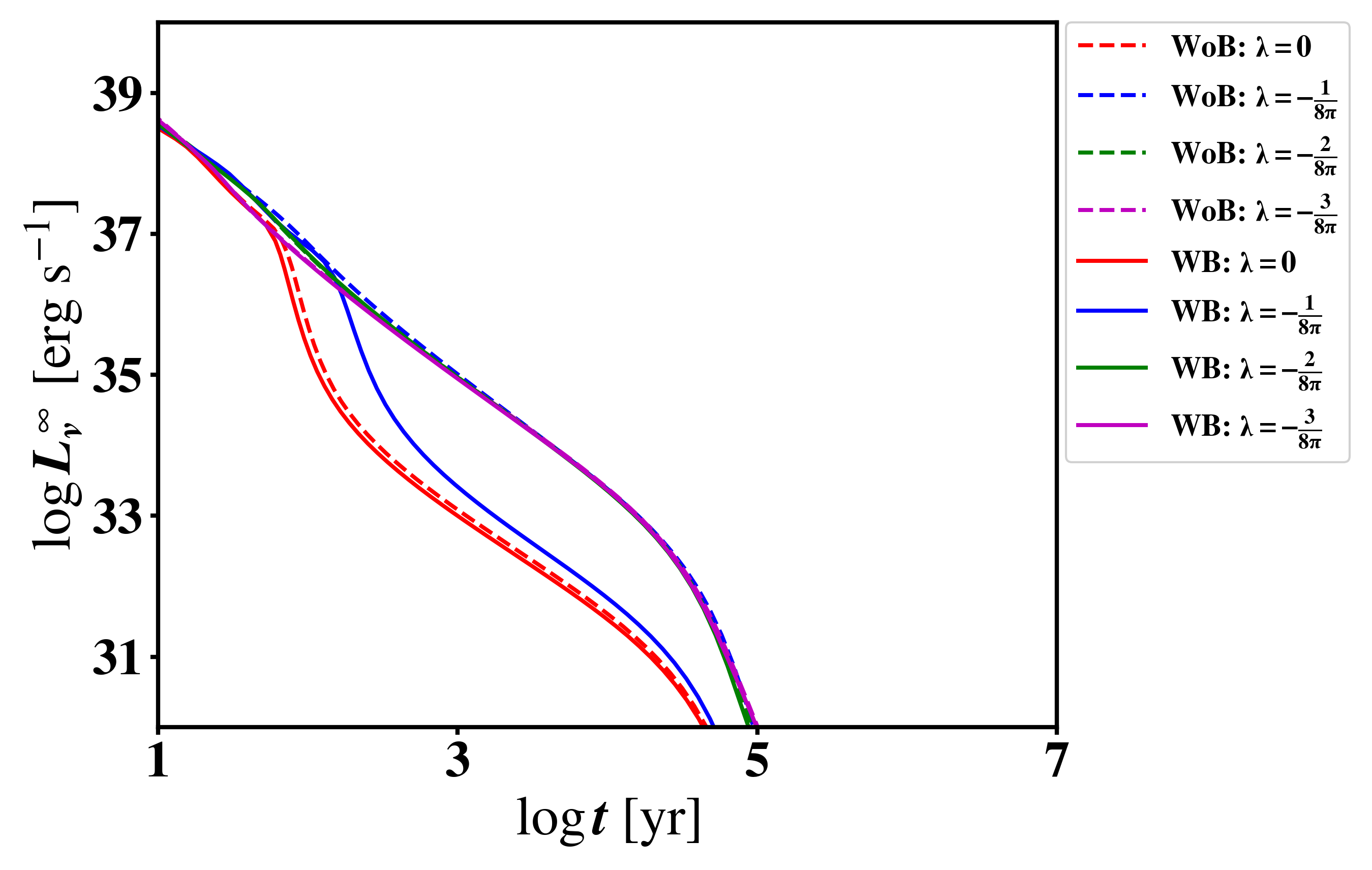}
    \caption{}
  \end{subfigure}

\caption{The variation of (a) red-shifted temperature ($T_s^\infty$), (b) red-shifted luminosity of photon ($L_\gamma^\infty$) and (c) red-shifted luminosity of neutrino ($L_\nu^\infty$) versus time for SLy EoS with and without magnetic field. Observed data have also been plotted here for comparison.}
\label{fig:sly}
\end{figure*}
The thermal evolution for magnetars and NSs in the framework of  $f(R, T)$ gravity using APR, FPS, and SLy EoSs is illustrated and discussed here. The inclusion of the $\lambda$ in $f(R, T)$ gravity demonstrates the matter curvature coupling, which is absent in the GR case. The chosen values of $\lambda$ are $\lambda = \frac{-1}{8\pi}$, $\lambda = \frac{-2}{8\pi}$, $\lambda = \frac{-3}{8\pi}$, which show a linear increase in coupling strength. The cooling behavior is characterized by the red-shifted surface temperature \(T_s^\infty(t)\), photon luminosity \(L_\gamma^\infty(t)\), and neutrino luminosity \(L_\nu^\infty(t)\). These quantities are depicted in Figs.~\ref{fig:apr}, \ref{fig:fps}, and \ref{fig:sly} for various values of the modified gravity parameter \(\lambda\), both with and without a central magnetic field of \(B_c = 10^{18}\,\mathrm{Gauss}\). Observational points for nearby isolated neutron stars are over-plotted for comparison, using data compiled in refs.~\cite {potekhin2020thermal,hambaryan2017compactness,hohle2012narrow,hohle2012continued,de2005polar,mignani2010optical,becker1997x,arumugasamy2018possible,mori2014broadband,walter1996discovery,mignani2013birthplace,ho2007magnetic,walter2010revisiting,potekhin2014atmospheres,sartore2012spectral,yoneyama2017discovery,kramer2003proper,ng2007origin}.\\

Figures~\ref{fig:apr}(a)–(c) illustrate the evolution of the red-shifted surface temperature ($T_s^\infty$), photon luminosity ($L_\gamma^\infty$), and neutrino luminosity ($L_\nu^\infty$) for different values of the parameter $\lambda$, both in the absence and presence of a magnetic field, using the APR EoS. Solid and dashed curves denote results with and without magnetic fields, respectively. At very early ages ($\sim 10$–$100$ yr), the cooling tracks for GR ($\lambda = 0$) and for $f(R,T)$ gravity with negative $\lambda$ nearly coincide. During this stage, cooling is dominated by core neutrino emission, rendering the influence of $\lambda$ negligible. The only noticeable deviation arises from the magnetic field: when a central magnetic field is included, the early-time $T_s^\infty$ and $L_\gamma^\infty$ curves lie slightly below those of the non-magnetic models.\\

Between roughly $10^2$ and slightly beyond $10^3$~yr, the influence of both $\lambda$ and the magnetic field on $T_s^\infty$ and $L_\gamma^\infty$ remains weak, indicating that the star is still in a neutrino-dominated cooling regime. At very early ages ($10$–$10^2$~yr), the red-shifted surface temperature and photon-luminosity curves for GR and modified gravity nearly coincide, while the effect of the magnetic field is already noticeable: for a central magnetic field of $B_c=10^{18}$~Gauss, the tracks of $T_s^\infty$ and $L_\gamma^\infty$ lie slightly below those obtained for $B_c=0$. As the evolution progresses into the $10^3$–$10^5$~yr interval, a systematic dependence on $\lambda$ emerges. More negative values of $\lambda$ shift both $T_s^\infty$ and $L_\gamma^\infty$ upward at a fixed age. In this same period, the magnetized curves lie slightly above the non-magnetized ones, reflecting the onset of the neutrino-to-photon cooling transition, during which reduced neutrino emissivity in $f(R, T)$ gravity and magnetic modifications to the envelope begin to influence observable quantities. At later times ($\gtrsim 10^5$~yr), the curves corresponding to different $\lambda$ values and magnetic field strengths converge, indicating that the effects of modified gravity and magnetic fields become negligible in the photon-dominated cooling era. The neutrino luminosity $L_\nu^\infty$ predicted by APR, shown in Fig.~\ref{fig:apr}(c), exhibits only small variations with $\lambda$ and $B$, except for minor differences at early times. It decreases smoothly over the lifetime and dominates over $L_\gamma^\infty$ up to $\sim 10^4$--$10^5$~yr, consistent with cooling controlled by modified Urca processes in a $1.4\,M_\odot$ NSs.\\

As compared to APR, FPS EoS produces faster cooling, as shown in Figs.~\ref{fig:fps}(a)–(c). The $T_s^\infty$ curves drop much more steeply at early times than in the APR case. In the first few tens of years, especially in the presence of the magnetic field, the surface temperature falls rapidly and the star becomes significantly cooler than in the APR and SLy (as shown in Fig.~\ref{fig:sly}) models at comparable ages. The photon luminosity $L_\gamma^\infty$ exhibits a characteristic sharp dip around $\sim 50$~yr, and then varies slowly up to $\sim 10^5$~yr, followed by a rapid decline afterwards. Varying $\lambda$ shifts the FPS temperature and photon-luminosity curves upward across the period of time; more negative $\lambda$ yields hotter and brighter tracks. However, even the warmest FPS models with $\lambda = -3/(8\pi)$ and $B_c \neq 0$ remain cooler and dimmer than the corresponding APR and SLy models. The neutrino luminosity is larger at early times and decays more rapidly than for APR, signaling a more efficient neutrino-cooling channel in the softer FPS core.\\

Physically, the soft FPS EoS leads to higher central densities and enhanced neutrino losses, so the internal energy is radiated away quickly and the star reaches low $T_s^\infty$ by $\sim 10^5$~yr. This rapid cooling cannot be compensated by the suppression of emissivity due to $\lambda < 0$ or by the presence of a strong magnetic field. As a result, FPS systematically underpredicts the observed surface temperatures and photon luminosities of middle-aged neutron stars, failing to comply with the observed data.\\

The SLy EoS exhibits a favorable cooling behavior in agreement with the observational data. The $T_s^\infty(t)$ in Fig.~\ref{fig:sly}(a) decreases relatively slowly at early times, steepens during the transition to photon cooling, and flattens again at late times. For $\lambda = -1/(8\pi)$, the presence of the magnetic field has a visible effect over a wide range of ages, as the magnetized curve maintains higher surface temperatures than the non-magnetized ones. When $\lambda$ shifts towards higher negative values, like ($-2/(8\pi)$ and $-3/(8\pi)$), the entire set of SLy $T_s^\infty$ curves is shifted upward, and the star remains systematically warmer, particularly between $10^3$ and $10^5$~yr. In contrast, in the GR limit  ($\lambda = 0$), the influence of the magnetic field on $T_s^\infty$ is small.\\

The photon luminosity $L_\gamma^\infty(t)$ in Fig.~\ref{fig:sly}(b) follows the same qualitative pattern as the temperature but displays even more pronounced differences between $\lambda$ values. For $\lambda = -2/(8\pi)$ and $-3/(8\pi)$, the magnetized SLy models stay significantly brighter during the neutrino–photon transition and throughout the photon-cooling era. The neutrino luminosity, shown in Fig.~\ref{fig:sly}(c), dominates up to $\sim 10^3$~yr and then falls, typically crossing the photon luminosity between $10^4$ and $10^5$~yr. For larger $|\lambda|$, the early-time $L_\nu^\infty$ is enhanced and its decay is more gradual, which leads to a smoother transition to photon-dominated cooling. The direct impact of the magnetic field on $L_\nu^\infty$ remains small. This behavior reflects the microphysics of a moderately stiff EoS. For SLy, a $1.4\,\rm M_\odot$ star stays in the slow-cooling (modified Urca) regime, avoiding strong direct Urca cooling. In this situation, a negative $f(R, T)$ coupling effectively reduces the net neutrino emissivity, while the magnetic field modifies heat transport and envelope blanketing. Together, these effects keep the star warmer and more luminous over an era.\\ 

We have compared model predictions with the measured red-shifted surface temperature of NSs and photon luminosity for PSR~J0538+2817~\cite{potekhin2020thermal,kramer2003proper,ng2007origin}, RX~J0720.4$-$3125~\cite{hambaryan2017compactness,potekhin2020thermal,hohle2012narrow,hohle2012continued}, PSR~B1055$-$52\cite{potekhin2020thermal,de2005polar,mignani2010optical}, PSR~B0656+14~\cite{becker1997x,de2005polar,arumugasamy2018possible,potekhin2020thermal}, Geminga~\cite{de2005polar,mori2014broadband}, and RX~J1856.5$-$3754~\cite{walter1996discovery,mignani2013birthplace,ho2007magnetic,walter2010revisiting,potekhin2014atmospheres,sartore2012spectral,yoneyama2017discovery} across their inferred ages. In general, models with negative $\lambda$ reproduce the data more consistently than pure GR; for $\lambda < 0$, the cooling tracks are uniformly warmer and brighter at a given age, without altering the epoch of the late-time decline. Among the EoSs, SLy gives the agreement that is well with the predicted values for $\lambda<0$ (with a modest magnetic uplift) passes through PSR~B0656+14 ($\sim10^{5}$ yr) and Geminga ($\gtrsim10^{5.5}$ yr), and brackets RX~J1856$-$3754. 
 APR EoS is broadly consistent, as at $\lambda<0$ raises the middle-age level, and a magnetic field adds only a small late-time offset. FPS EoS improves with $\lambda<0$, but remains systematically cooler/dimmer, and yields the weakest agreement. Early times ($\lesssim10^{3}$ yr) are largely insensitive to $\lambda$ and $B_c$. The cleanest discrimination arises near the neutrino-to-photon transition ($\sim10^{4}$-
$10^{6}$ yr). In summary, the observations favor $f(R, T)$ with SLy+$\lambda<0$ (optionally magnetized) as a reasonable match to the data, while APR+$\lambda<0$ is also acceptable. In contrast, our prediction using the FPS EoS is less favorable compared to the data.\\

 It is evident from the observed data that SLy EoS demonstrates a reasonable agreement between the observed and predicted values of \(T_S^\infty \) and $L_\gamma^\infty(t)$, for higher absolute non-zero \( \lambda \) values varying from \( \lambda = -\frac{2}{8\pi}, -\frac{3}{8\pi}\). This indicates that, in the presence of a magnetic field, modified gravity performs well for the SLy EoS and yields favorable results.\\

 The parameter \( \lambda \) modifies the cooling rate by altering the internal thermodynamic properties of neutron stars. Figures \ref{fig:sly}(a) illustrate that larger absolute values of \( \lambda \) correspond to a slower cooling rate for SLy EoS in the case of a magnetic field. This behavior suggests that stronger deviations from standard gravity (larger \( |\lambda| \)) lead to modifications in the neutrino emissivity and energy transport processes, thus flattening the evolution curves \(T_s^\infty\) and delaying the cooling process.\\


\section{Conclusions and Future Outlook}
\label{conc}

In this work, we have investigated the thermal evolution of neutron stars within the linear $f(R, T)$ gravity model, $f(R, T) = R + 2\lambda\kappa T$, incorporating the effects of a strong central magnetic field and three representative nucleonic equations of state (APR, FPS, and SLy). Using magnetized TOV solutions as input to the NSCool framework, we computed the red-shifted surface temperature, photon luminosity, and neutrino luminosity for a $1.4\,\mathrm{M}_\odot$ neutron star, and compared the resulting cooling tracks with observational data from isolated neutron stars.

\begin{itemize}
\item Our analysis reveals that negative values of the matter–curvature coupling $\lambda$ systematically slow the cooling, yielding warmer and more luminous stars at a given age than in GR ($\lambda=0$), while preserving the overall morphology of the cooling curves. A strong internal magnetic field ($B_c = 10^{18},\mathrm{Gauss}$) further modifies the evolution by altering heat transport through the envelope and enhancing the late-time surface temperature and photon luminosity. However, the magnitude of this effect is strongly EoS dependent.

\item By examining the cooling evolution at fixed mass across APR, FPS, and SLy, we isolate EoS-specific trends from mass–radius degeneracies. The three models exhibit a robust ordering during the neutrino-cooling era (APR hotter, FPS cooler, SLy intermediate), while both negative $\lambda$ and magnetization act as additive warmers once the photon era is reached. The combined effect is particularly effective for the SLy EoS, where the cooling tracks match benchmark observations at $t \sim 10^{5}-10^{6} \mathrm{yr}$ without invoking fine-tuned envelope compositions or artificial suppression of neutrino emissivities.

\item The influence of $\lambda<0$ manifests as a coherent, time-dependent modification across all three EoSs. Though the impact of  $\lambda$ shows a marginal effect in NSs cooling and luminosity corresponding to the APR EoS. Early neutrino cooling dominated by core microphysics remains nearly unaltered, but as the star approaches the photon-cooling era, the reduction in neutrino emissivity inherent to $f(R, T)$ gravity increases $T_s^\infty$ and $L_\gamma^\infty$ relative to GR. This results in a delayed and smoother transition from neutrino to photon dominance, an effect most visible for $t \sim 10^{5}-10^{6} \mathrm{yr}$, where observational constraints are strongest.

\item Magnetization ($B_c \sim 10^{18},\mathrm{Gauss}$) acts in the same direction as the $f(R, T)$ coupling in the photon era, elevating the late-time surface temperature and photon luminosity while leaving the early-time, neutrino-dominated decline largely intact. Thus, negative $\lambda$ and a strong internal field operate as additive warmers after the neutrino phase, shifting the cooling trajectories upward without distorting their qualitative shapes, which remain controlled principally by the EoS.

\item Comparison with observed surface temperatures and photon luminosities of six standard isolated neutron stars—PSR J0538+2817, PSR B0656+14, PSR B1055$-$52, Geminga, RX J0720.4$-$3125, and RX J1856.5$-$3754 shows that this combined effect provides a compelling match. In the photon-cooling epoch, the observational data cluster within a narrow region at $t \sim 10^{5}-10^{6} \mathrm{yr}$, through which the SLy sequences with $\lambda<0$ and $B_c\neq0$ pass naturally. APR and FPS, while still influenced by the exact mechanisms, require comparatively larger deviations to match these data.

\item  Taken together, these findings demonstrate that at fixed stellar mass, the EoS governs the overall cooling temperature\rm and the onset of photon dominance, while negative $\lambda$ in $f(R, T)$ gravity and strong internal magnetic fields provide late-time warming that improves agreement with photon-era observations without compromising the neutrino-era behavior. Among the configurations explored, the SLy EoS combined with $\lambda < 0$ and $B_c \neq 0$ yields the most consistent qualitative and quantitative agreement across the complete set of cooling diagnostics $T_s^{\infty}(t)$, $L_\gamma^{\infty}(t)$, and $L_\nu^{\infty}(t)$ for a $1.4 M_\odot$ neutron star.
\end{itemize}

Future extensions may explore other stellar masses, more realistic magnetic-field geometries, additional microphysics such as hyperons or quark matter, and more general $f(R, T)$ functional forms to tighten constraints on modified gravity further using neutron-star thermal evolution.

\section {Acknowledgments}
C.R. acknowledges financial support from the Department of Science and Technology, New Delhi, through a DST–INSPIRE Fellowship. C.R. is grateful to Birla Institute of Technology and Science, Pilani (Pilani Campus, Rajasthan 333031) for research facilities and administrative support. We thank S. Banik and K. Nobleson for insightful discussions on $f(R)$ gravity.
 
\bibliographystyle{apsrev4-2} 
\bibliography{references}

@article{dexheimer2017magnetic,
  title={Magnetic field distribution in strongly magnetized neutron stars},
  author={Dexheimer, V and Franzon, B and Gomes, RO and Farias, RLS and Avancini, SS},
  journal={Astronomische Nachrichten},
  volume={338},
  number={9-10},
  pages={1052--1055},
  year={2017},
  publisher={Wiley Online Library}
}

@article{mahapatra2024neutron,
  title = {Neutron Stars in Modified $f(R, T)$ Gravity Framework with $\mathcal{O}(T, T^2)$ Terms},
  author = {Mahapatra, Premachand and Das, Prasanta Kumar},
  journal = {arXiv preprint},
  eprint = {2401.01321},
  archivePrefix = {arXiv},
  year = {2024}
}

@article{carvalho2017stellar,
  title={Stellar equilibrium configurations of white dwarfs in the f (R, T) gravity},
  author={Carvalho, GA and Lobato, RV and Moraes, PHRS and Arba{\~n}il, Jos{\'e} DV and Otoniel, E and Marinho, RM and Malheiro, M},
  journal={The European Physical Journal C},
  volume={77},
  pages={1--8},
  year={2017},
  publisher={Springer}
}

@article{pretel2021neutron,
  title={Neutron stars in f (R, T) gravity with conserved energy-momentum tensor: Hydrostatic equilibrium and asteroseismology},
  author={Pretel, Juan MZ and Jor{\'a}s, Sergio E and Reis, Ribamar RR and Arba{\~n}il, Jos{\'e} DV},
  journal={Journal of Cosmology and Astroparticle Physics},
  volume={2021},
  number={08},
  pages={055},
  year={2021},
  publisher={IOP Publishing}
}

@article{harko2011f,
  title={f (R, T) gravity},
  author={Harko, Tiberiu and Lobo, Francisco SN and Nojiri, Shin’ichi and Odintsov, Sergei D},
  journal={Physical Review D—Particles, Fields, Gravitation, and Cosmology},
  volume={84},
  number={2},
  pages={024020},
  year={2011},
  publisher={APS}
}

@article{sotiriou2010f,
  title={f (R) theories of gravity},
  author={Sotiriou, Thomas P and Faraoni, Valerio},
  journal={Reviews of Modern Physics},
  volume={82},
  number={1},
  pages={451--497},
  year={2010},
  publisher={APS}
}

@article{de2010f,
  title={f (R) theories},
  author={De Felice, Antonio and Tsujikawa, Shinji},
  journal={Living Reviews in Relativity},
  volume={13},
  number={1},
  pages={1--161},
  year={2010},
  publisher={Springer}
}

@article{tolman1939static,
  title={Static solutions of Einstein's field equations for spheres of fluid},
  author={Tolman, Richard C},
  journal={Physical Review},
  volume={55},
  number={4},
  pages={364},
  year={1939},
  publisher={APS}
}

@article{oppenheimer1939massive,
  title={On massive neutron cores},
  author={Oppenheimer, J Robert and Volkoff, George M},
  journal={Physical Review},
  volume={55},
  number={4},
  pages={374},
  year={1939},
  publisher={APS}
}

@article{yadav2024thermal,
  title={Thermal evolution and axion emission properties of strongly magnetized neutron stars},
  author={Yadav, Shubham and Mishra, M and Sarkar, Tapomoy Guha and Singh, Captain R},
  journal={The European Physical Journal C},
  volume={84},
  number={3},
  pages={225},
  year={2024},
  publisher={Springer}
}

@article{wijnands2017cooling,
  title={Cooling of accretion-heated neutron stars},
  author={Wijnands, Rudy and Degenaar, Nathalie and Page, Dany},
  journal={Journal of Astrophysics and Astronomy},
  volume={38},
  pages={1--16},
  year={2017},
  publisher={Springer}
}

@article{ofengeim2014neutrino,
  title={Neutrino-pair bremsstrahlung in a neutron star crust},
  author={Ofengeim, DD and Kaminker, AD and Yakovlev, DG},
  journal={Europhysics Letters},
  volume={108},
  number={3},
  pages={31002},
  year={2014},
  publisher={IOP Publishing}
}

@article{prakash1994rapid,
  title={Rapid cooling of neutron stars},
  author={Prakash, M},
  journal={Physics Reports},
  volume={242},
  number={4-6},
  pages={297--312},
  year={1994},
  publisher={Elsevier}
}

@article{iwamoto1995neutrino,
  title={Neutrino magnetic moment and neutron star cooling},
  author={Iwamoto, Naoki and Qin, Letao and Fukugita, Masataka and Tsuruta, Sachiko},
  journal={Physical Review D},
  volume={51},
  number={2},
  pages={348},
  year={1995},
  publisher={APS}
}

@article{page2016nscool,
  title={NSCool: Neutron star cooling code},
  author={Page, Dany},
  journal={Astrophysics Source Code Library},
  pages={ascl--1609},
  year={2016}
}

@article{page2006cooling,
  title={The cooling of compact stars},
  author={Page, Dany and Geppert, Ulrich and Weber, Fridolin},
  journal={Nuclear Physics A},
  volume={777},
  pages={497--530},
  year={2006},
  publisher={Elsevier}
}

@article{brown1988strangeness,
  title={Strangeness condensation and cooling of neutron stars},
  author={Brown, GE and Kubodera, K and Page, Dany and Pizzochero, Pierre},
  journal={Physical Review D},
  volume={37},
  number={8},
  pages={2042},
  year={1988},
  publisher={APS}
}

@article{beznogov2023standard,
  title={Standard cooling of rapidly rotating isolated neutron stars in 2D},
  author={Beznogov, Mikhail V and Novak, J{\'e}r{\^o}me and Page, Dany and Raduta, Adriana R},
  journal={The Astrophysical Journal},
  volume={942},
  number={2},
  pages={72},
  year={2023},
  publisher={IOP Publishing}
}

@article{yakovlev2004neutron,
  title={Neutron star cooling: theoretical aspects and observational constraints},
  author={Yakovlev, DG and Gnedin, Oleg Y and Kaminker, AD and Levenfish, KP and Potekhin, Alexander Y},
  journal={Advances in Space Research},
  volume={33},
  number={4},
  pages={523--530},
  year={2004},
  publisher={Elsevier}
}

@article{potekhin2015neutron,
  title={Neutron stars—cooling and transport},
  author={Potekhin, Alexander Y and Pons, Jos{\'e} A and Page, Dany},
  journal={Space Science Reviews},
  volume={191},
  pages={239--291},
  year={2015},
  publisher={Springer}
}

@article{buschmann2022upper,
  title={Upper limit on the QCD axion mass from isolated neutron star cooling},
  author={Buschmann, Malte and Dessert, Christopher and Foster, Joshua W and Long, Andrew J and Safdi, Benjamin R},
  journal={Physical review letters},
  volume={128},
  number={9},
  pages={091102},
  year={2022},
  publisher={APS}
}

@article{page2009neutrino,
  title={Neutrino emission from cooper pairs and minimal cooling of neutron stars},
  author={Page, Dany and Lattimer, James M and Prakash, Madappa and Steiner, Andrew W},
  journal={The Astrophysical Journal},
  volume={707},
  number={2},
  pages={1131},
  year={2009},
  publisher={IOP Publishing}
}

@article{leinson2000neutrino,
  title={Neutrino emission due to Cooper pairing of protons in cooling neutron stars: Collective effects},
  author={Leinson, LB},
  journal={Physics Letters B},
  volume={473},
  number={3-4},
  pages={318--323},
  year={2000},
  publisher={Elsevier}
}

@article{yakovlev2001neutrino,
  title={Neutrino emission from neutron stars},
  author={Yakovlev, DG and Kaminker, AD and Gnedin, Oleg Y and Haensel, P},
  journal={Physics Reports},
  volume={354},
  number={1-2},
  pages={1--155},
  year={2001},
  publisher={Elsevier}
}

@article{yakovlev1999cooling,
  title={Cooling of neutron stars and superfluidity in their cores},
  author={Yakovlev, Dmitrii G and Levenfish, Kseniya P and Shibanov, Yurii A},
  journal={Physics-Uspekhi},
  volume={42},
  number={8},
  pages={737},
  year={1999},
  publisher={IOP Publishing}
}

@article{baiko1998direct,
  title={Direct urca process in strong magnetic fields and neutron star cooling},
  author={Baiko, DA and Yakovlev, DG},
  journal={arXiv preprint astro-ph/9812071},
  year={1998}
}

@article{kaminker1999neutrino,
  title={Neutrino emission due to electron bremsstrahlung in superfluid neutron-star cores},
  author={Kaminker, AD and Haensel, P},
  journal={arXiv preprint astro-ph/9908249},
  year={1999}
}

@article{gusakov2005cooling,
  title={The cooling of Akmal—Pandharipande—Ravenhall neutron star models},
  author={Gusakov, ME and Kaminker, AD and Yakovlev, Dima G and Gnedin, Oleg Y},
  journal={Monthly Notices of the Royal Astronomical Society},
  volume={363},
  number={2},
  pages={555--562},
  year={2005},
  publisher={Blackwell Science Ltd Oxford, UK}
}

@article{akmal1998equation,
  title={Equation of state of nucleon matter and neutron star structure},
  author={Akmal, A and Pandharipande, VR and Ravenhall, DG},
  journal={Physical Review C},
  volume={58},
  number={3},
  pages={1804},
  year={1998},
  publisher={APS}
}

@article{schneider2019akmal,
  title={Akmal-Pandharipande-Ravenhall equation of state for simulations of supernovae, neutron stars, and binary mergers},
  author={Schneider, Andre S and Constantinou, C and Muccioli, Brian and Prakash, M},
  journal={Physical Review C},
  volume={100},
  number={2},
  pages={025803},
  year={2019},
  publisher={APS}
}

@article{flowers1976neutrino,
  title={Neutrino pair emission from finite-temperature neutron superfluid and the cooling of young neutron stars},
  author={Flowers, Elliott and Ruderman, Malvin and Sutherland, Peter},
  journal={Astrophysical Journal, vol. 205, Apr. 15, 1976, pt. 1, p. 541-544.},
  volume={205},
  pages={541--544},
  year={1976}
}

@article{broderick2000equation,
  title={The equation of state of neutron star matter in strong magnetic fields},
  author={Broderick, A and Prakash, M and Lattimer, JM},
  journal={The Astrophysical Journal},
  volume={537},
  number={1},
  pages={351},
  year={2000},
  publisher={IOP Publishing}
}

@article{ng2007origin,
  title={The Origin and Motion of PSR J0538+ 2817 in S147},
  author={Ng, C-Y and Romani, Roger W and Brisken, Walter F and Chatterjee, Shami and Kramer, Michael},
  journal={The Astrophysical Journal},
  volume={654},
  number={1},
  pages={487},
  year={2007},
  publisher={IOP Publishing}
}

@article{beznogov2021heat,
  title={Heat blanketing envelopes of neutron stars},
  author={Beznogov, MV and Potekhin, AY and Yakovlev, DG},
  journal={Physics Reports},
  volume={919},
  pages={1--68},
  year={2021},
  publisher={Elsevier}
}

@article{potekhin2001thermal,
  title={Thermal structure and cooling of neutron stars with magnetized envelopes},
  author={Potekhin, A Yu and Yakovlev, DG},
  journal={Astronomy \& Astrophysics},
  volume={374},
  number={1},
  pages={213--226},
  year={2001},
  publisher={EDP Sciences}
}

@article{chugunov2007thermal,
  title={Thermal conductivity of ions in a neutron star envelope},
  author={Chugunov, AI and Haensel, P},
  journal={Monthly Notices of the Royal Astronomical Society},
  volume={381},
  number={3},
  pages={1143--1153},
  year={2007},
  publisher={Blackwell Publishing Ltd Oxford, UK}
}

@article{bocquet1995rotating,
  title={Rotating neutron star models with magnetic field},
  author={Bocquet, M and Bonazzola, S and Gourgoulhon, E and Novak, J},
  journal={arXiv preprint gr-qc/9503044},
  year={1995}
}

@article{chatterjee2015consistent,
  title={Consistent neutron star models with magnetic-field-dependent equations of state},
  author={Chatterjee, Debarati and Elghozi, Thomas and Novak, Jerome and Oertel, Micaela},
  journal={Monthly Notices of the Royal Astronomical Society},
  volume={447},
  number={4},
  pages={3785--3796},
  year={2015},
  publisher={Oxford University Press}
}

@article{moraes2016stellar,
  title={Stellar equilibrium configurations of compact stars in f (R, T) theory of gravity},
  author={Moraes, PHRS and Arba{\~n}il, Jos{\'e} DV and Malheiro, M},
  journal={Journal of Cosmology and Astroparticle Physics},
  volume={2016},
  number={06},
  pages={005},
  year={2016},
  publisher={IOP Publishing}
}

@article{nobleson2022comparison,
  title={Comparison of perturbative and non-perturbative methods in f (R) gravity},
  author={Nobleson, K and Ali, Amna and Banik, Sarmistha},
  journal={The European Physical Journal C},
  volume={82},
  number={1},
  pages={32},
  year={2022},
  publisher={Springer}
}

@article{nava2023probing,
  title={Probing strong field f (r) gravity and ultradense matter with the structure and thermal evolution of neutron stars},
  author={Nava-Callejas, Martin and Page, Dany and Beznogov, Mikhail V},
  journal={Physical Review D},
  volume={107},
  number={10},
  pages={104057},
  year={2023},
  publisher={APS}
}

@article{dohi2021neutron,
  title={Neutron star cooling in modified gravity theories},
  author={Dohi, Akira and Kase, Ryotaro and Kimura, Rampei and Yamamoto, Kazuhiro and Hashimoto, Masa-aki},
  journal={Progress of Theoretical and Experimental Physics},
  volume={2021},
  number={9},
  pages={093E01},
  year={2021},
  publisher={Oxford University Press}
}

@article{beznogov2016cooling,
  title={Cooling of neutron stars with diffusive envelopes},
  author={Beznogov, MV and Fortin, M and Haensel, P and Yakovlev, DG and Zdunik, JL},
  journal={Monthly Notices of the Royal Astronomical Society},
  volume={463},
  number={2},
  pages={1307--1313},
  year={2016},
  publisher={Oxford University Press}
}

@article{bhattacharjee2024white,
  title={White dwarf cooling in f (R, T) gravity},
  author={Bhattacharjee, Snehasish},
  journal={International Journal of Modern Physics A},
  volume={39},
  number={05n06},
  pages={2450026},
  year={2024},
  publisher={World Scientific}
}

@article{kalita2022cooling,
  title={Cooling process of white dwarf stars in Palatini f (R) gravity},
  author={Kalita, Surajit and Sarmah, Lupamudra and Wojnar, Aneta},
  journal={Universe},
  volume={8},
  number={12},
  pages={647},
  year={2022},
  publisher={MDPI}
}

@article{kalita2023metric,
  title={Metric-affine effects in crystallization processes of white dwarfs},
  author={Kalita, Surajit and Sarmah, Lupamudra and Wojnar, Aneta},
  journal={Physical Review D},
  volume={107},
  number={4},
  pages={044072},
  year={2023},
  publisher={APS}
}

@article{douchin2001unified,
  title={A unified equation of state of dense matter and neutron star structure},
  author={Douchin, F and Haensel, P},
  journal={Astronomy \& Astrophysics},
  volume={380},
  number={1},
  pages={151--167},
  year={2001},
  publisher={EDP Sciences}
}

@article{friedman1981hot,
  title={Hot and cold, nuclear and neutron matter},
  author={Friedman, B and Pandharipande, VR},
  journal={Nuclear Physics A},
  volume={361},
  number={2},
  pages={502--520},
  year={1981},
  publisher={Elsevier}
}

@book{haug2004elementary,
  title={The elementary process of bremsstrahlung},
  author={Haug, Eberhard and Nakel, Werner},
  volume={73},
  year={2004},
  publisher={World Scientific}
}

@article{kramer2003proper,
  title={The proper motion, age, and initial spin period of PSR J0538+ 2817 in S147},
  author={Kramer, Michael and Lyne, AG and Hobbs, G and L{\"o}hmer, O and Carr, P and Jordan, C and Wolszczan, A},
  journal={The Astrophysical Journal},
  volume={593},
  number={1},
  pages={L31},
  year={2003},
  publisher={IOP Publishing}
}

@article{becker1997x,
  title={The X-ray luminosity of rotation-powered neutron stars},
  author={Becker, Werner and Tr{\"u}mper, Joachim},
  journal={arXiv preprint astro-ph/9708169},
  year={1997}
}

@article{de2005polar,
  title={On the polar caps of the three musketeers},
  author={De Luca, A and Caraveo, PA and Mereghetti, S and Negroni, M and Bignami, GF},
  journal={The Astrophysical Journal},
  volume={623},
  number={2},
  pages={1051},
  year={2005},
  publisher={IOP Publishing}
}

@article{arumugasamy2018possible,
  title={Possible phase-dependent absorption feature in the x-ray spectrum of the middle-aged PSR J0659+ 1414},
  author={Arumugasamy, Prakash and Kargaltsev, Oleg and Posselt, Bettina and Pavlov, George G and Hare, Jeremy},
  journal={The Astrophysical Journal},
  volume={869},
  number={2},
  pages={97},
  year={2018},
  publisher={IOP Publishing}
}

@article{mignani2010optical,
  title={OPTICAL--ULTRAVIOLET SPECTRUM AND PROPER MOTION OF THE MIDDLE-AGED PULSAR B1055- 52},
  author={Mignani, RP and Pavlov, GG and Kargaltsev, O},
  journal={The Astrophysical Journal},
  volume={720},
  number={2},
  pages={1635},
  year={2010},
  publisher={IOP Publishing}
}

@article{mori2014broadband,
  title={A broadband X-ray study of the Geminga pulsar with NuSTAR and XMM-Newton},
  author={Mori, Kaya and Gotthelf, Eric V and Dufour, Francois and Kaspi, Victoria M and Halpern, Jules P and Beloborodov, Andrei M and An, Hongjun and Bachetti, Matteo and Boggs, Steven E and Christensen, Finn E and others},
  journal={The Astrophysical Journal},
  volume={793},
  number={2},
  pages={88},
  year={2014},
  publisher={IOP Publishing}
}

@article{hambaryan2017compactness,
  title={The compactness of the isolated neutron star RX J0720. 4- 3125},
  author={Hambaryan, V and Suleimanov, V and Haberl, F and Schwope, AD and Neuh{\"a}user, R and Hohle, M and Werner, K},
  journal={Astronomy \& Astrophysics},
  volume={601},
  pages={A108},
  year={2017},
  publisher={Edp Sciences}
}

@article{sartore2012spectral,
  title={Spectral monitoring of RX J1856. 5-3754 with XMM-Newton-Analysis of EPIC-pn data},
  author={Sartore, N and Tiengo, A and Mereghetti, S and De Luca, A and Turolla, Roberto and Haberl, F},
  journal={Astronomy \& Astrophysics},
  volume={541},
  pages={A66},
  year={2012},
  publisher={EDP Sciences}
}

@article{potekhin2020thermal,
  title={Thermal luminosities of cooling neutron stars},
  author={Potekhin, AY and Zyuzin, DA and Yakovlev, DG and Beznogov, MV and Shibanov, Yu A},
  journal={Monthly Notices of the Royal Astronomical Society},
  volume={496},
  number={4},
  pages={5052--5071},
  year={2020},
  publisher={Oxford University Press}
}

@article{rathod2025structural,
  title={Structural Properties of Magnetized Neutron Stars under f (R, T) Gravity Framework},
  author={Rathod, Charul and Mishra, M and Das, Prasanta Kumar},
  journal={arXiv preprint arXiv:2511.00425},
  year={2025}
}

@article{yadav2024x,
  title={X-ray emission spectrum for axion--photon conversion in magnetospheres of strongly magnetized neutron stars},
  author={Yadav, Shubham and Mishra, Madhukar and Sarkar, Tapomoy Guha},
  journal={The European Physical Journal C},
  volume={84},
  number={7},
  pages={687},
  year={2024},
  publisher={Springer}
}

@article{gudmundsson1983structure,
  title={Structure of neutron star envelopes},
  author={Gudmundsson, Einar H and Pethick, CJ and Epstein, Richard I},
  journal={Astrophysical Journal, Part 1 (ISSN 0004-637X), vol. 272, Sept. 1, 1983, p. 286-300.},
  volume={272},
  pages={286--300},
  year={1983}
}

@article{nomoto1987cooling,
  title={Cooling of neutron stars-Effects of the finite time scale of thermal conduction},
  author={Nomoto, Kenichi and Tsuruta, Sachiko},
  journal={Astrophysical Journal, Part 1 (ISSN 0004-637X), vol. 312, Jan. 15, 1987, p. 711-726.},
  volume={312},
  pages={711--726},
  year={1987}
}

@article{friman1979neutrino,
  title={Neutrino emissivities of neutron stars},
  author={Friman, BL and Maxwell, OV},
  journal={Astrophysical Journal, Part 1, vol. 232, Sept. 1, 1979, p. 541-557.},
  volume={232},
  pages={541--557},
  year={1979}
}

@article{walter1996discovery,
  title={Discovery of a nearby isolated neutron star},
  author={Walter, Frederick M and Wolk, Scott J and Neuh{\"a}user, Ralph},
  journal={Nature},
  volume={379},
  number={6562},
  pages={233--235},
  year={1996},
  publisher={Nature Publishing Group UK London}
}

@article{mignani2013birthplace,
  title={The birthplace and age of the isolated neutron star RX J1856. 5-3754},
  author={Mignani, RP and Putte, D Vande and Cropper, M and Turolla, Roberto and Zane, S and Pellizza, LJ and Bignone, LA and Sartore, N and Treves, Aldo},
  journal={Monthly Notices of the Royal Astronomical Society},
  volume={429},
  number={4},
  pages={3517--3521},
  year={2013},
  publisher={Oxford University Press}
}

@article{ho2007magnetic,
  title={Magnetic hydrogen atmosphere models and the neutron star RX J1856. 5--3754},
  author={Ho, Wynn CG and Kaplan, David L and Chang, Philip and Van Adelsberg, Matthew and Potekhin, Alexander Y},
  journal={Monthly Notices of the Royal Astronomical Society},
  volume={375},
  number={3},
  pages={821--830},
  year={2007},
  publisher={Blackwell Publishing Ltd Oxford, UK}
}

@article{walter2010revisiting,
  title={Revisiting the parallax of the isolated neutron star RX J185635- 3754 using HST/ACS imaging},
  author={Walter, FM and Eisenbei{\ss}, T and Lattimer, JM and Kim, B and Hambaryan, V and Neuh{\"a}user, R},
  journal={The Astrophysical Journal},
  volume={724},
  number={1},
  pages={669},
  year={2010},
  publisher={IOP Publishing}
}

@article{potekhin2014atmospheres,
  title={Atmospheres and radiating surfaces of neutron stars},
  author={Potekhin, Alexander Y},
  journal={Physics-Uspekhi},
  volume={57},
  number={8},
  pages={735},
  year={2014},
  publisher={IOP Publishing}
}

@article{yoneyama2017discovery,
  title={Discovery of a keV-X-ray excess in RX J1856. 5--3754},
  author={Yoneyama, Tomokage and Hayashida, Kiyoshi and Nakajima, Hiroshi and Inoue, Shota and Tsunemi, Hiroshi},
  journal={Publications of the Astronomical Society of Japan},
  volume={69},
  number={3},
  pages={50},
  year={2017},
  publisher={Oxford University Press}
}

@article{hohle2012narrow,
  title={Narrow absorption features in the co-added XMM--Newton RGS spectra of isolated neutron stars},
  author={Hohle, Markus M and Haberl, Frank and Vink, Jacco and de Vries, Cor P and Neuh{\"a}user, Ralph},
  journal={Monthly Notices of the Royal Astronomical Society},
  volume={419},
  number={2},
  pages={1525--1536},
  year={2012},
  publisher={Blackwell Publishing Ltd Oxford, UK}
}

@article{hohle2012continued,
  title={The continued spectral and temporal evolution of RX J0720. 4- 3125},
  author={Hohle, Markus M and Haberl, Frank and Vink, Jacco and de Vries, Cor P and Turolla, Roberto and Zane, Silvia and M{\'e}ndez, Mariano},
  journal={Monthly Notices of the Royal Astronomical Society},
  volume={423},
  number={2},
  pages={1194--1199},
  year={2012},
  publisher={Blackwell Publishing Ltd Oxford, UK}
}

\end{document}